\documentclass[11pt]{article}
\usepackage{graphicx,cite,amsmath}

\renewcommand{\appendix}
        {
        \par
        \setcounter{section}{0}
        \setcounter{subsection}{0}
        \gdef\afterthesectionpunctdefault{:}
        \gdef\thesection{{Appendix \Alph{section}}}
        \renewcommand{\theequation}{\Alph{section}\arabic{equation}}
        \setcounter{equation}{0}
        }
\def\lsim{\hbox{\lower .8ex\hbox{$\, \buildrel < \over \sim\,$}}}
\def\gsim{\hbox{\lower .8ex\hbox{$\, \buildrel > \over \sim\,$}}}
\newcommand{\APPROX}[1]{                
   {{\raisebox{-.3cm}{$\textstyle\simeq$}} \atop {\scriptstyle{#1}}}}
\def\ds{\displaystyle}

\def\ep{{\mbox{\large e}}}
\def\sss{\scriptscriptstyle}

\newfont{\ensmathquatorze}{msbm10 scaled 1400}
\newfont{\ensmathonze}{msbm10 scaled 1100}
\newfont{\ensmathdix}{msbm10}
\newfont{\ensmathneuf}{msbm10 scaled 833}
\newfont{\ensmathhuit}{msbm10 scaled 694}
\newfam\ensmathfam                        
\textfont\ensmathfam=\ensmathonze        
\scriptfont\ensmathfam=\ensmathdix       
\scriptscriptfont\ensmathfam=\ensmathhuit
\def\ensmf{\fam\ensmathfam\ensmathonze}         

\def\NN{{\ensmf N}}                 
\def\RR{{\ensmf R}}                 

\begin{document}

\begin{center}

{\huge Dark soliton past a finite-size obstacle}

\vspace{1.0cm}

{\Large Nicolas Bilas and Nicolas Pavloff}
\end{center}

\begin{center}
\noindent Laboratoire de Physique Th\'eorique
et Mod\`eles Statistiques\footnotemark,\\
\vspace{0.1 cm}
Universit\'e Paris Sud, b\^at. 100, F-91405 Orsay Cedex, France \\
\vspace{0.1 cm}

\vspace{2 cm}
{\bf Abstract}
\end{center}

We consider the collision of a dark soliton with an obstacle in a quasi-one-
dimensional Bose condensate. We show that in many respects the soliton behaves
as an effective classical particle of mass twice the mass of a bare particle,
evolving in an effective potential which is a convolution of the actual
potential describing the obstacle. Radiative effects beyond this approximation
are also taken into account. The emitted waves are shown to form two counterpropagating wave packets, both moving at the speed of sound. We determine, at leading order, the total amount of radiation emitted during the collision and compute the acceleration of the soliton due to the collisional process. It is found that the radiative process is quenched when the velocity of the soliton reaches the velocity of sound in the system.

\vspace{3 cm}
\begin{math}
\footnotetext[1]{Unit\'e Mixte de Recherche de l'Universit\'e Paris XI et du
CNRS (UMR 8626).}
\end{math}

\noindent PACS numbers:
\vspace{0.5 cm}

\noindent 03.75.-b Matter waves\hfill\break
\noindent 05.60.Gg Quantum transport\hfill\break
\noindent 42.65.Tg Optical solitons; nonlinear guided waves\hfill\break
\newpage

\section{Introduction}

One of the many interesting aspects of the physics of Bose-Einstein condensation
of ultracold atomic vapors is to open opportunities of studying
mesoscopiclike phenomena in new types of setups. The 
advances in the production and propagation of Bose-Einstein condensates in more
and more elaborate waveguides (magnetic or optical, microfabricated or not
\cite{guides}) opens up the prospect of studying a rich variety of quantum
transport phenomena for these intrinsically phase-coherent, finite-sized
systems. In particular it has been possible to study quantum interference
effects \cite{Shi04}, Bloch oscillations and Landau-Zener tunneling \cite{BO},
Josephson junctions \cite{JJ}, and superfluidity \cite{super}.

\

Pushing further the analogy in transport properties of mesoscopic systems and
Bose-condensed vapors, one notices that, whereas in mesoscopic physics
interaction effects are often difficult to understand, in Bose-Einstein
condensates they are more easily accessible to theoretical description and
have the advantage of covering a wide range of regimes, ranging from almost
noninteracting atom lasers to strongly correlated systems. Along this line, the
existence of nonlinearity in the wave equation, resulting in the existence of
bright \cite{BS} a dark \cite{DS} solitons, appears as a natural -- and rather
simply understood -- consequence of interaction on transport phenomena of
quasi-one-dimensional Bose-condensed systems.

\

In the present work we address the problem of transport of a dark soliton in a
quasi-one-dimensional Bose-Einstein condensate. More precisely, we consider a
guided Bose-Einstein condensate and theoretically study the propagation of a
dark soliton encountering an obstacle on its way. In the appropriate limit
[see Eq. (\ref{s1.0}) below] the system is described by a one-dimensional nonlinear Schr\"odinger equation. This equation admits bright and dark solitonic
solutions, depending on the sign of the interparticle interaction. The
obstacle is modeled via an external potential, and this could correspond to
different physical realizations, such as a heavy impurity, a (red or blue)
detuned laser beam crossing the atomic beam, a bend, a twist, or a constriction
in the shape of the guide.

\

A soliton under the influence of a perturbation (here, the obstacle) sees its
shape and velocity modified and may also radiate energy (see, e.g., Ref.
\cite{Kee77}). Despite their mutual dependence, these two phenomena are not
easily treated on the same theoretical footing. The evolution of the soliton's
parameters is typically studied within the adiabatic approximation (see
Ref.~\cite{Kiv89} and references therein), whereas radiative effects are not so
easily described, because their influence on the soliton's parameters only
appears at second order in perturbation theory (see the discussion in Section
\ref{rad_ener}). However, it has been possible to treat both phenomena
concomitantly in the case of bright solitons
\cite{Kee77,Kiv89,Kar79,Kau78,Kau90}. Concerning dark solitons, several
studies of adiabatic dynamics have appeared
\cite{Kiv94,Kiv95,Kon97,Bus01,Fra02,Bra03,Kon04}, but until recently radiative
effects have been treated mainly numerically \cite{Bur97,Par03,Par04}.

  In the present paper we study the dynamics of a dark soliton via
perturbation theory. This method, based on the
theory of linear partial differential equations, has been established in
the case of the nonlinear Schr\"odinger equation with repulsive interaction in
Refs. \cite{Chen98,Hua99} (see also the earlier attempt \cite{Kon94}). 
Although our first interest lies in the physics
of guided Bose-Einstein condensates, the method employed and the results 
displayed also apply to optical waveguides described by a one-dimensional (1D)
nonlinear defocussing Schr\"odinger equation.

\

The paper is organized as follows. In Section \ref{model} we present the basic
ingredients of the model and the resulting equation governing the time
evolution of the condensate wave function. In the framework of perturbation
theory we then derive the equations determining the dynamics of the soliton
and of the radiated part (Section \ref{perturbation}). The results are
analysed in Section \ref{analysis}. We show that one can devise a quite
successful approximation that we denote as ``effective potential theory,''
where the soliton is assimilated to a classical particle of mass twice the
mass of a bare particle, evolving in an effective potential (Section
\ref{eff}). The agreement of this approximation with the results of the
adiabatic approximation is verified even for a fine quantity such as the
position shift induced on the trajectory of the soliton by the obstacle
(Section \ref{num}). We then consider in Section \ref{bfwp} the radiated part
and show that it is formed of backward- and forward-emitted phonons, which form
two counterpropagating wavepackets mooving at the speed of sound. In the
limit of large soliton's velocity we furthermore obtain in Section
\ref{rad_ener} an analytical expression for the total amount of radiation
emitted by the soliton during the collision. In addition we show that (within
our leading-order evaluation) a soliton reaching the velocity of sound does
not radiate, and we propose a physical interpretation for this phenomenon.
Finally we present our conclusions in Section \ref{conclusion}. Some technical
points are given in the Appendixes. In \ref{modespropres} we recall the main
properties of the spectrum of the operator governing the wave dynamics of the
system around the solitonic solution. In \ref{lagrangien} we briefly present
the Lagrangian approach for deriving the dynamics of the parameters of a dark
soliton. In \ref{calcul-ener-rad} we show how to compute some integrals
involved in the evaluation of the total amount of radiation emitted by the
soliton.

\section{The model}\label{model}

We consider a condensate confined in a guide of axis $z$ and denote by
$n(z,t)$ the 1D density of the system. The condensate is formed by atoms of
mass $m$ which interact  via a two-body potential characterized by its 3D
s-wave scattering length $a_{sc}$. We consider the case of a repulsive
effective interaction-i.e., $a_{sc}>0$. The condensate is confined in the
transverse direction by an harmonic potential of pulsation $\omega_\perp$. The
transverse confinement is characterized by the harmonic oscillator length
$a_\perp=(\hbar/m\omega_\perp)^{1/2}$.

With $n_{\sss 1D}$ denoting a typical order of magnitude of $n(z,t)$, we restrict
ourselves to a density range such that
\begin{equation}\label{s1.0}
(a_{sc}/a_\perp)^2\ll n_{\sss 1D}\,a_{sc} \ll 1\; .
\end{equation}
This regime has been called ``1D mean field'' in Ref. \cite{Men02}. In this
range the wave function of the condensate can be factorized in a transverse
and longitudinal part \cite{Jac98,Ols98,Leb01}. The transverse wave function
is Gaussian (this is ensured by the condition $n_{\sss 1D}\,a_{sc} \ll 1$) and
the longitudinal one, denoted by $\psi(z,t)$ [such that
$n(z,t)=|\psi(z,t)|^2$], satisfies an effective 1D Gross-Pitaevskii equation
(see, e.g., \cite{Jac98,Ols98,Leb01}):
\begin{equation}\label{s1.1}
-\frac{\hbar^2}{2\,m}\psi_{z z} + 
\Big\{U(z) + 2\hbar\omega_\perp\,a_{sc}|\psi|^2\Big\}\,\psi =
 i\hbar\,\psi_t \; .
\end{equation} 

\

In Eq. (\ref{s1.1}), $U(z)$ represents the effect of the obstacle. We restrict
ourselves to the case of localized obstacle such that
$\lim_{z\to\pm\infty}U(z)=0$. Hence, we can consider that the stationary
solutions of Eq.(\ref{s1.1}) have at infinity an asymptotic density unperturbed
by the obstacle. Besides, considering solutions without current at infinity,
we impose the following form to the the stationary solutions:
\begin{equation}\label{s1.2}
\psi_{\mbox{\tiny sta}}(z,t)=f(z)\,\exp[-i\mu\,t/\hbar] \; ,\quad\mbox{with}\quad
\lim_{z\to\pm\infty} f(z)=\sqrt{n_\infty} \; ,
\end{equation}
\noindent where $n_\infty$ is the 1D density far from the obstacle and 
$\mu=2\hbar\omega_\perp\,a_{sc}n_\infty$ the chemical potential \cite{r2}.

\

	We note here that in Eq.(\ref{s1.0}) we have discarded very low densities in
order to prevent the system from getting in the Tonks-Girardeau regime where the
mean-field picture implicit in Eq.~(\ref{s1.1}) breaks down
\cite{Ols98,Pet00}. This can be intuitively understood as follows: it is
natural to assume that the Gross-Pitaevskii scheme is valid-i.e., that the
system can be described by a collective order parameter $\psi$-only if the
interparticle distance (of order $n_{\infty}^{-1}$) is much smaller than the
minimum distance $\xi$ over which $\psi$ can significantly vary [$\xi$ is the
healing length, defined by $\xi=\hbar/(m\,\mu)^{1/2}=a_\perp/(2
a_{sc}n_\infty)^{1/2}$]. The condition $n_\infty^{-1}\ll\xi$ then imposes us to
consider the regime $n_\infty\,a_{sc}\gg (a_{sc}/a_\perp)^2$ to which, from Eq.
(\ref{s1.0}), we restrict our study. If one considers, for instance, $^{87}$Rb
or $^{23}$Na atoms in a guide with a transverse confinement characterized by
$\omega_\perp=2\pi\times 500$ Hz, the ratio $a_{sc}/a_\perp$ is roughly of
order $10^{-2}$ and the restriction (\ref{s1.0}) still allows the density to
vary over four orders of magnitude.

\

In all the following we use dimensionless quantities: the energies are
expressed in units of $\mu$, the lengths in units of $\xi$, and the time in
units of $\hbar/\mu$. $\psi$ is also rescaled by a factor $n_\infty^{-1/2}$;
this corresponds to expressing the linear density in units of the 
density at infinity, $n_\infty$. We keep the same notation $z$, $t$, $U(z)$, and
$\psi(z,t)$ for the rescaled quantities. Equation (\ref{s1.1}) now reads
\begin{equation}\label{s1.3}
-\frac{1}{2}\psi_{z z} + 
\Big\{U(z) + |\psi|^2\Big\}\,\psi =
 i\,\psi_t \; .
\end{equation} 
From Eq.(\ref{s1.2}), the stationary solutions of Eq.(\ref{s1.3}) are of type
$f(z)\exp[-i\,t]$, $f$ being real, and a solution of
\begin{equation}\label{s1.4}
-\frac{1}{2}f_{z z} + 
\Big\{U(z) + f^2-1\Big\}\,f = 0\; ,
\end{equation}
\noindent with the asymptotic condition $\lim_{z\to\pm\infty} f(z)=1$.

The method we will expose is quite general and applies to a broad range of
potentials $U(z)$, but for concreteness we will often display the explicit
solutions of the problem in the case of a pointlike obstacle, where
$U(z)=\lambda\,\delta(z)$; $\lambda>0$ ($<0$) corresponds to a repulsive
(attractive) obstacle. For such an obstacle, the solution of Eq.(\ref{s1.4}) is
\begin{equation}\label{s1.5}
f(z)=\left\{
\begin{array}{lcc}
\tanh(|z|+a) & \mbox{if} & \lambda>0 \\
\coth(|z|+a) & \mbox{if} & \lambda<0 \\
\end{array}\right.
\qquad \mbox{with}\qquad a=\frac{1}{2} \,
\sinh^{-1}\left(\frac{2}{|\lambda|}\right) \; .
\end{equation}

In section 4 we will concentrate on perturbative aspects of the problem and
consider the case of a weak potential $U(z)$. For a pointlike obstacle,
this corresponds to the limit $|\lambda|\ll 1$. In this case
$\sinh^{-1}(2/|\lambda|)\simeq\ln(4/|\lambda|)$ and Eqs.~(\ref{s1.5})
simplify to
\begin{equation}\label{s1.6}
f(z)\simeq 1-\frac{\lambda}{2}\exp\{-2|z|\}\; .
\end{equation}
In the general case, one can design a simple treatment \cite{Hak97,Leb01}
valid for any weak potential $U(z)$ leading after linearization of
Eq.~(\ref{s1.4}) to the perturbative result $f(z)=1+\delta f(z)$ with
\begin{equation}\label{s1.7}
\delta f(z)\simeq-\frac{1}{2}\int_{-\infty}^{+\infty}\!\!\!dy \;
U(y)\exp\{-2|z-y|\} \; ,
\end{equation}
\noindent of which Eq.(\ref{s1.6}) is a particular case. In section 4 it will
reveal convenient to rewrite Eq.(\ref{s1.7}) in an other way: denoting by
$\hat{U}(q)=\int_\RR dz\,U(z)\exp(-i q z)$ the Fourier transform of $U(z)$, one
may equivalently express $\delta f$ defined in Eq.(\ref{s1.7}) as
\begin{equation}\label{s1.7bis}
\delta f(z)\simeq-2\, \int_{-\infty}^{+\infty} \frac{dq}{2\pi} 
\,\frac{\hat{U}(q)}{4+q^2}\,\exp\{i q z\}\; .
\end{equation}
The stationary solutions of the problem being defined, let us now turn to the
main subject of the present work and consider the case of time-dependent
solutions corresponding to a dark soliton pro\-pa\-ga\-ting in the system. The
soliton will appear as a distortion of the stationary background, and it is
here very natural to follow the approach of Frantzeskakis {\it et al.}
\cite{Fra02} who write the wave function of the system as a product:
\begin{equation}\label{s1.8}
\psi(z,t)=\phi(z,t)f(z)\exp(-i\,t)\; .
\end{equation}
$\phi(z,t)$ in Eq.~(\ref{s1.8}) accounts for the deformation of the
stationary background $f(z)\exp(-i\,t)$ caused by the motion of a soliton in
the system. From Eq.~(\ref{s1.3}) we see that the unknown field $\phi(z,t)$
is a solution of the following equation:
\begin{equation}\label{s1.9}
i\phi_t+\frac{1}{2}\phi_{z z}-\Big\{|\phi|^2-1\Big\}\phi=R[\phi]
\; ,
\end{equation}
\noindent where 
\begin{equation}\label{s1.10}
R[\phi]=-\phi_z\,\frac{f_z}{f}+(f^2-1)(|\phi|^2-1)\phi \; .
\end{equation}

Far from the obstacle, $f(z)=1$ and thus $R[\phi]=0$. In this case, the motion
of a dark soliton in the system is described by the usual solitonic solution
of the defocussing nonlinear Schr\"odinger equation~\cite{Kiv98}
\begin{equation}\label{s1.11}
\phi(z,t)=\Phi(z-Vt-b,\theta) \; ,
\end{equation}
\noindent where
\begin{equation}\label{s1.12}
\Phi(x,\theta)=\chi(x,\theta)+i\,V \; ,\quad\mbox{with}\quad
\chi(x,\theta)=\cos\theta\,\tanh(x\,\cos\theta) \quad\mbox{and}\quad
\sin\theta=V\; .
\end{equation}
Equations (\ref{s1.11}) and (\ref{s1.12}) describe a dark soliton consisting in a
density trough located at position $V\,t+b$ at time $t$. The phase change
across the soliton is $2\theta-\pi$. The choice of the parameter $\theta$ in
$[0,\pi/2]$ corresponds to a soliton moving from left to right with a
velocity $V=\sin\theta\in[0,1]$. Note that a dark soliton has a velocity
always lower than unity (which, in our rescaled units, is the velocity of sound
\cite{r1}). When $\theta=0$, the soliton is standing and its minimum density
is zero; it is referred to as a black soliton. When $\theta\ne 0$ one speaks of
a gray soliton. We display in Fig. 1 the density profile and the
phase of the wave function $\psi(z,t)$ [see Eq.~(\ref{s1.8})] describing a
soliton incident with velocity $V=0.4$ on a repulsive point-like obstacle
characterized by $\lambda=0.5$.

\

\begin{minipage}{8 cm}
\includegraphics*[width=7cm]{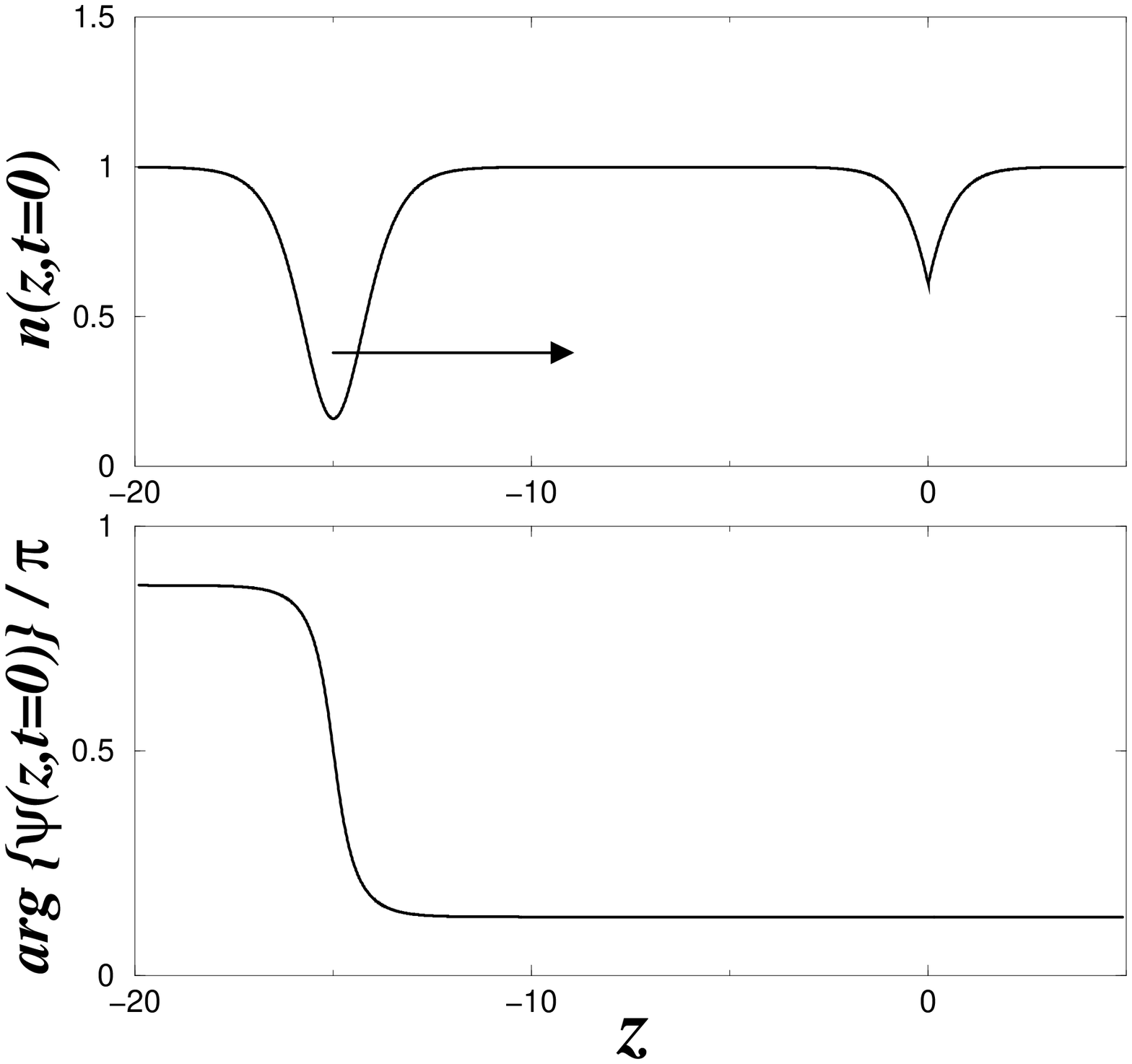}
\end{minipage}
\begin{minipage}{7 cm}
Figure 1: {\sl Upper plot: density profile of a dark soliton incident with
velocity $V=0.4$ on a point-like repulsive obstacle $U(z)=\lambda\,\delta(z)$
(with $\lambda=0.5$). The arrow represents the direction of propagation of the
soliton. Lower plot: phase of the wave function $\psi(z,t=0)$ describing the
system. Across the soliton the phase of the wave function changes from
$\pi-\theta$ to $\theta$ (with $V=\sin\theta$).}
\end{minipage}

\section{Perturbation theory}\label{perturbation}

In the following we will set up the basis for a systematic perturbative
expansion, and for properly identifying the orders of perturbation at which
the expansion is done, it is customary to introduce an artificial
multiplicative parameter $\epsilon$ in the potential of the obstacle
(otherwise of arbitrary form). We will see in the present section (and justify
on physical grounds in the next one) that for an obstacle
characterized by $\epsilon\, U(z)$, the condition of small perturbation reads
$V^2\gg \epsilon\, U$. Since the soliton velocity is always lower than unity
(which is the speed of sound in our dimensionless units), this condition
implies $\epsilon\, U\ll 1$; i.e., Eqs.~(\ref{s1.7},\ref{s1.7bis}) hold.

At initial times the soliton is unperturbed and described as in the previous
section by $\phi(z,t)=\Phi(z-Vt,\theta_0)$ [$\Phi$ is defined in
Eq.~(\ref{s1.12})]; i.e., one considers a soliton incident from left infinity
with velocity $V=\sin\theta_0$. The more important effect of the obstacle on
the soliton is a modification of its shape; i.e., the parameters
characterizing the soliton will become time dependent in the vicinity of the
obstacle. Perturbation at next order describe the emission of radiations.
One thus looks for solutions of Eq.(\ref{s1.9}) of the form 
\begin{equation}\label{p1}
\phi(z,t)=\phi_{\mbox{\tiny{sol}}}(z,\bar{z}(t),\theta(t))+\delta\phi(z,t)\; ,
\end{equation}
\noindent where
\begin{equation}\label{p1a}
\phi_{\mbox{\tiny{sol}}}(z,\bar{z}(t),\theta(t))=\Phi(z-\bar{z}(t),\theta(t))\; ,
\end{equation}
describes a soliton which is characterized by the two parameters $\bar{z}(t)$
(decribing the center of the soliton) and $\theta(t)$ (describing the phase shift
accross the soliton). $\delta\phi$ describes additional radiative components:
\begin{equation}\label{p1b}
\delta\phi(z,t)=\epsilon\,\phi_1(z,t)+\epsilon^2\,\phi_2(z,t)+...\; .
\end{equation}
Equations (\ref{p1})-(\ref{p1b}) form the grounds of a secular
perturbation theory where the time dependence of the parameters of the soliton
permits to avoid the grow of secular perturbation in $\delta\phi$ (see, e.g.,
the discussion in Ref.~\cite{Kee77}). 

It is more appropriate to define $\delta\phi$ in Eq.(\ref{p1}) and the $\phi_i$'s in Eq.(\ref{p1b}) as functions of $z-\bar{z}(t)$ than as functions of $z$. To this end, we define $x=z-\bar{z}(t)$ and choose to work with $x$ and $t$ as independent parameters rather than $z$ and $t$. This corresponds to the transformation
\begin{equation}\label{p2}
\partial_z \; , \; \partial_t \quad\to\quad
\partial_x \; , \; \partial_t-\dot{\bar{z}}\,\partial_x\; .
\end{equation}
Furthermore, in order to take into account the slow time dependence of the
parameters of the soliton, it is customary to introduce multiple time scales:
$t_n=\epsilon^n\,t$ ($n\in\NN$). A time-dependent function could, for instance,
depend on $t$ via $t_1$, indicating a weak time dependence (a $t_2$ dependence
being related to an even weaker time dependence and a $t_0$ dependence to a
``normal'' time dependence). Generically, time-dependent quantities will be
considered as functions of all the $t_n$'s, with
\begin{equation}\label{p3}
\partial_t=\partial_{t_0}+\epsilon\,\partial_{t_1}+\epsilon^2\,\partial_{t_2}
+...\; .
\end{equation}
In the following we will make an expansion at order $\epsilon$ and
it will suffice to consider only the fast time $t_0$ and the first slow time
$t_1$. The soliton's parameters $\theta$ and $\bar{z}$ are considered as
functions $\theta(t_1)$ and $\bar{z}(t_0,t_1)$ \cite{trem}.

Putting everything together, we see that, at order $\epsilon$, Eqs.
(\ref{p1})-(\ref{p1b}) read explicitly
\begin{equation}\label{p4}
\phi(z,t)=\Phi(x,\theta(t_1))+\epsilon\,\phi_1(x,t_0,t_1)\; ,
\quad\mbox{with}\quad x=z-\bar{z}(t_0,t_1) \; .
\end{equation}
 Equation(\ref{s1.9}) is now rewritten taking the transformations (\ref{p2}) and
(\ref{p3}) into account, with an expansion at order $\epsilon$. To this end, we
have to take into account that $R[\phi]$ defined in Eq.(\ref{s1.10}) is a small
quantity and can be written at first order in $\epsilon$
as 
\begin{equation}\label{p5}
R[\phi]\simeq -\partial_x\Phi(x,\theta) \,
\big[\partial_z\delta f(z) +2\,\Phi(x,\theta)\delta f(z)\big]
\equiv\epsilon\,{\cal R}(x,z)
\; ,
\end{equation}
\noindent where $z=x+\bar{z}(t_0,t_1)$ 
and $\delta f(z)$ is defined as in Eqs.(\ref{s1.7}) and (\ref{s1.7bis}),
with an extra multiplicative factor $\epsilon$ in $U$ which has been written
explicitly in the definition of ${\cal R}$ on the right-hand-side (RHS) of Eq.(\ref{p5}).

We are now ready to expand Eq.(\ref{s1.9}) in successive orders in
$\epsilon$. The leading order reads
\begin{equation}\label{p6}
-\frac{1}{2}\,\Phi_{xx}+i\bar{z}_{t_0}\,\Phi_x+(|\Phi|^2-1)\Phi
=0\; ,
\end{equation}
implying that
\begin{equation}\label{p7}
\bar{z}_{t_0}=\sin\theta\; ,
\end{equation}
whence $\bar{z}$ can be written as
\begin{equation}\label{p8}
\bar{z}=t_0\,\sin\theta+\tilde{z}(t_1) \; ,
\end{equation}
\noindent where $\tilde{z}(t_1)$ is a still unknown function \cite{att}.
At next order in $\epsilon$ one obtains
\begin{equation}\label{p9}
i\partial_{t_0}\phi_1=\big[-\frac{1}{2}\partial^2_x+i\sin\theta\,
\partial_x+2|\Phi|^2-1\big]\phi_1+\Phi^2\,\phi_1^*
+{\cal R}
-i\theta_{t_1}\,\Phi_\theta
+i\bar{z}_{t_1}\,\Phi_x \; .
\end{equation}
Equation (\ref{p9}) can be rewritten as
\begin{equation}\label{p10}
i\partial_{t_0}|\phi_1\rangle = {\cal H}\,|\phi_1\rangle
+\sigma_3\,|{\cal R}\,\rangle
+i\bar{z}_{t_1}|\omega_e\rangle
-\frac{\theta_{t_1}}{\cos\theta}\,|\Omega_e\rangle
\; ,
\end{equation}
\noindent where $|\phi_1\rangle=(\phi_1 , \phi_1^*)^{\sss
T}$, $|{\cal R}\,\rangle=({\cal R} , {\cal R}^*)^{\sss T}$, $\sigma_3$ is the
third Pauli matrice, and
\begin{equation}\label{s3.10}
{\cal H}=\left(
\begin{array}{cc}
-\frac{1}{2}\partial_x^2+i\sin\theta\,\partial_x+2|\Phi|^2-1 & \Phi^2 \\
-\Phi^{*\,2} & \frac{1}{2}\partial_x^2+i\sin\theta\,\partial_x-2|\Phi|^2+1 \\
\end{array}
\right) \; . 
\end{equation}
${\cal H}$ is not diagonalizable, but can be put in a Jordan form in a manner
similar to what has been done for the attractive nonlinear Schr\"odinger
equation \cite{Wei85}. Its eigenfunctions and eigenvalues are presented
in \ref{modespropres}. In particular, $|\omega_e\rangle$ and
$|\Omega_e\rangle$ appearing in Eq.(\ref{p10}) belong to the generalized null
space of ${\cal H}$; they  verify ${\cal H}|\omega_e\rangle=0$
and ${\cal H}|\Omega_e\rangle=\cos^2\theta\,|\omega_e\rangle$. As well as its null
space, ${\cal H}$ has two continuous branches of excitations which we denote
by its ``phonon spectrum.'' The corresponding eigenfunctions are denoted by 
$|\Xi_q^\pm\rangle$ with $q\in\RR$ (see \ref{modespropres}).

It is physically intuitive that $|\phi_1\rangle$ corresponding to the radiated
part should be expanded over the phonon part of the spectrum of ${\cal
H}$:
\begin{equation}\label{s3.13}
|\phi_1\rangle = \sum_{\alpha=\pm}
\int_{-\infty}^{+\infty}\!\!\!dq \;
C_q^\alpha(t_0,t_1)\;|\Xi_q^\alpha\rangle \; .
\end{equation}
A more technical argument for limiting the expansion (\ref{s3.13}) to the
phonon components of the spectrum of ${\cal H}$ is the following:  one might
think that a greater generality could be achieved by allowing $|\phi_1\rangle$
to have also components on $|\omega_e\rangle$ and $|\Omega_e\rangle$, for
instance. However, exactly as in the case of the bright soliton \cite{Kau90},
these components can (and should) be imposed to remain zero for avoiding
the appearance of secular terms in the evolution of the soliton's parameters.

\subsection{Evolution of the parameters of the soliton}

Applying $\langle\omega_e|\sigma_3$ and $\langle\Omega_e|\sigma_3$ onto 
(\ref{p10}) and using the orthogonality relations (\ref{b6}), one obtains the
equations of evolution of the parameters of the
soliton:
\begin{eqnarray}\label{s3.14}
4\,\theta_{t_1}\,\cos^2\theta=-\langle\omega_e|{\cal R}\,\rangle
& = & -2\;\mbox{Re}\;\left\{\int_{-\infty}^{+\infty}\!\!\!dx\,
\Phi_x \, {\cal R}^*(x,x+\bar{z}) \right\} \nonumber \\
& = & \frac{2}{\epsilon}\;\mbox{Re}\;\left\{\int_{-\infty}^{+\infty}\!\!\!dz\,
R^*[\phi_{\mbox{\tiny{sol}}}]\;\partial_{\bar{z}}
\phi_{\mbox{\tiny{sol}}} \right\}\; ,
\end{eqnarray}
\noindent and
\begin{eqnarray}\label{s3.15}
4\,\bar{z}_{t_1}\,\cos^2\theta
=  \frac{1}{i\cos\theta}\;
\langle\Omega_e|{\cal R}\,\rangle 
& = & -2 \;
\mbox{Re}\;\left\{\int_{-\infty}^{+\infty}\!\!\!dx \;
\Phi_\theta\,
{\cal R}^*(x,x+\bar{z})\right\} \nonumber \\
& = &
-\frac{2}{\epsilon}\;\mbox{Re}\;\left\{\int_{-\infty}^{+\infty}\!\!\!dz \;
\,R^*[\phi_{\mbox{\tiny{sol}}}]\;
\partial_\theta \phi_{\mbox{\tiny{sol}}}\right\} \; .
\end{eqnarray}

The set of equations (\ref{p7}),(\ref{s3.14}) and (\ref{s3.15}) describe the time
evolution of the soliton's parameter. The same equations are obtained via
adiabatic approximation which is a simpler variationnal approximation where
radiative effects are neglected [see \ref{lagrangien}, Eqs.
(\ref{s2.6.1}) and (\ref{s2.6.2})]. This is evident in the case of Eq. (\ref{s3.14})
which is the slow time analogous to Eq. (\ref{s2.6.1}) (since
$\dot{\theta}=\epsilon\,\theta_{t_1}$). In a similar way, the prescription
(\ref{p3}) indicates that $\dot{\bar{z}} =
\bar{z}_{t_0}+\epsilon\,\bar{z}_{t_1}$; combining Eqs. (\ref{p7}) and
(\ref{s3.15}), one sees that the equations of evolution of $\bar{z}$ obtained
in the present section correspond to the multiple-time expansion of
Eq.(\ref{s2.6.2}). As a side result of this exact correspondance of the time
evolution of the soliton's parameters, we obtain here that, as in the
adiabatic approach, $\sin[\theta(t_1\to\pm\infty)]=V$ (see the discussion at
the end of \ref{lagrangien}) and the quantity $\tilde{z}$ appearing in Eq.
(\ref{p8}) is identical when $t_1\to+\infty$ to the one defined in Eq.
(\ref{s2.16.b}).

\

A technical remark is in order here. One can notice that in Eq.(\ref{p1}) we did
not consider the most general variational form for the solitonic component of
the wave function. We could have let its global phase depend on time, for
instance, and this would have given in Eq.(\ref{p10}) a contribution along
$|\omega_o\rangle$ ($|\omega_o\rangle$ is defined in \ref{modespropres}).
Similarly, a more general variational ansatz could also have been used in
\ref{lagrangien}. The important point is that if the soliton's parameters are
chosen within the same variational space, their time evolution is described --
in the adiabatic and perturbative approach -- by the same equations. Besides,
the radiative term $\phi_1$ having in all cases to be restricted to the phonon
part of the spectrum, its time evolution is not (at least in the limit $V^2\gg
\epsilon\,U$; see below) affected by the specific choice of variational
parameters used for describing the soliton.

\subsection{Radiated part}\label{radpart}

  The time evolution of the radiative component $|\phi_1\rangle$ is obtained in
a manner similar to what is done for the soliton's parameters. Projecting
Eq.~(\ref{p10}) onto the phonon eigenfunctions of ${\cal H}$ by applying
$\langle\Xi_q^\alpha|\sigma_3$ yields
\begin{equation}\label{s3.16}
i \, {\cal N}_q^\alpha \,\partial_{t_0} C_q^\alpha =
{\cal N}_q^\alpha \,\varepsilon_q^\alpha\, C_q^\alpha+
\langle\Xi_q^\alpha |{\cal R}\rangle \; ,
\end{equation}
\noindent where $\alpha=\pm$. $\varepsilon_q^\alpha$ in Eq.(\ref{s3.16}) is the
eigenvalue of ${\cal H}$ associated with $|\Xi_q^\alpha\rangle$ [see Eq.(\ref{bb})]
\begin{equation}\label{s3.16ter}
\varepsilon_q^\alpha=
q\,\left(-\sin\theta+\alpha\sqrt{\frac{q^2}{4}+1}\,\right)\; ,
\end{equation}
\noindent and ${\cal N}_q^\alpha$ is a normalization factor [see Eqs.(\ref{b4}) 
and (\ref{b5})]. In deriving Eq.(\ref{s3.16}), we have taken into account that
the eigenfunctions $|\Xi_q^\alpha\rangle$ depend on $t$ only through the
slow time $t_1$ (via $\sin\theta$). The same holds for 
$\varepsilon_q^\alpha$ and ${\cal N}_q^\alpha$. Thus, writing
\begin{equation}\label{s3.16bis}
C_q^\alpha(t_0,t_1)=D_q^\alpha(t_0,t_1)\,\exp\{-i\,\varepsilon_q^\alpha t_0\}\; ,
\end{equation}
\noindent one has, at the same order of approximation as Eq.(\ref{s3.16}),
\begin{equation}\label{s3.16.b}
\partial_{t_0}D_q^\alpha=\frac{1}{i\,{\cal N}_q^\alpha} \,
\langle\Xi_q^\alpha |{\cal R}\rangle \; .
\end{equation}
In integrating Eq.~(\ref{s3.16.b}) we can choose between two equivalent
strategies. The first (and difficult) one is to solve this equation taking
into account that $t_1=\epsilon\,t_0$ and that $\theta$ and $\bar{z}$ have the
time dependence specified by Eqs.(\ref{p7}),(\ref{s3.14}) and (\ref{s3.15}). The
second one is to integrate this equation considering $t_0$ and $t_1$ as
independent variables. In this case, the $t_1$ dependence of $\theta$ and
$\bar{z}$ will not matter and the $t_0$ dependence of $\bar{z}$ will be
specified by Eq.(\ref{p8}). According to this second method one obtains
\begin{equation}\label{s3.17}
D_q^\alpha(t_0,t_1)=
\frac{1}{i\,{\cal N}^\alpha_q}\,\int_{-\infty}^{t_0}\!\!dt_0'\,
\ep^{\ds i\varepsilon_q^\alpha t_0'}\;\langle\Xi_q^\alpha|{\cal R}\rangle
+\tilde{D}_q^\alpha(t_1) \; ,
\end{equation}
\noindent where $\tilde{D}_q^\alpha$ is an unknown function of $t_1$
[verifying $\tilde{D}_q^\alpha(t_1\to-\infty)=0$] which could be determined
by pushing the perturbative expansion to next order in $\epsilon$. In the
following we will simply neglect this term. This is legitimate in the limit
where all the $t_1$-dependent terms are nearly constant-i.e., to the limit
where the parameters of the soliton are very weakly affected by the obstacle.
We will see in the next section that this limit is reached when $\epsilon\,
U\ll V^2$.

\

Of most interest to us is the total amount of radiation emitted by the
soliton. For the determination of this quantity 
we need the explicit expression of
Eq.(\ref{s3.17}) at large times. In the limit $t_0$ and $t_1\to+\infty$, 
Eq.~(\ref{s3.17}) (without the $\tilde{D}_q^\alpha$ term) reads explicitly
\begin{equation}\label{s3.18}
D_q^\alpha(+\infty)=\frac{1}{i\,{\cal N}^\alpha_q}\,\int_{\RR^2}\!\!\!
dx\,dt_0'\;\ep^{\ds i\varepsilon_q^\alpha\,t_0'}\;
\Big[u_q^{\alpha\, *}(x)\,{\cal R}(x,x+\bar{z}(t_0',+\infty))+
v_q^{\alpha\, *}(x)\,{\cal R}^*(x,\bar{z}(t_0',+\infty))\Big]\; ,
\end{equation}
\noindent where the functions $u_q^{\alpha}(x)$ and $v_q^{\alpha}(x)$ are the
explicit components of $|\Xi_q^\alpha\rangle$ defined in \ref{modespropres}
[Eq.~(\ref{b3})]. Note that the $t_1$-dependent parameters in Eq.(\ref{s3.18})
have been given their asymptotic value. In particular,
$\sin\theta(t_1\to+\infty) = \sin\theta_0 = V$, and according to Eq.
(\ref{p8}) one has here $\bar{z}(t_0',+\infty)=V\,t_0' + \tilde{z}(+\infty)$.
The integration along $t_0'$ in Eq.(\ref{s3.18}) can be computed easily using the
expression (\ref{p5}) for ${\cal R}$ and Eq.(\ref{s1.7bis}) for $\delta f$,
leading to
\begin{eqnarray}\label{s3.19}
D_q^\alpha(+\infty)& = & \frac{4}{i\,V\,{\cal N}^\alpha_q}\,
\frac{\hat{U}^*(\frac{\varepsilon_q^\alpha}{V})}
{4+(\frac{\varepsilon_q^\alpha}{V})^2}\, 
\ep^{ -i\varepsilon_q^\alpha \tilde{z}(+\infty)/V}\,
\times\nonumber \\
& & \int_{\RR}\!\!\!
dx\;\ep^{ -i\varepsilon_q^\alpha x/V}\,\partial_x\Phi\,
\left[u_q^{\alpha\, *}(x)
\left(\Phi(x)-\frac{i \varepsilon_q^\alpha}{2\,V}\right)+
v_q^{\alpha\, *}(x)
\left(\Phi(x)^*-\frac{i \varepsilon_q^\alpha}{2\,V}\right)
\right]\; .
\end{eqnarray}
A long but straightforward computation gives the final result
\begin{equation}\label{s3.20}
D_q^\alpha(+\infty)=- \frac{1}{16\,V^3}\,
\frac{q}{\sqrt{1+q^2/4}}\,
\frac{\hat{U}^*(\varepsilon_q^\alpha/V)\,
\ep^{ -i\varepsilon_q^\alpha \tilde{z}(+\infty)/V}\,}
{\sinh\left(\frac{\ds\pi\,q\,\sqrt{1+q^2/4}}{\ds 2\,V\,\sqrt{1-V^2}}\right)}
\; .
\end{equation}
In this formula the term $\tilde{z}(+\infty)$ can be obtained through the
numerical determination of $\bar{z}(t)$. We indicate in section \ref{num}
different approximation schemes allowing one to obtain an analytical evaluation of
this term [Eq. (\ref{s2.16}) and below]. 
>From expression (\ref{s3.20}) we see that the radiation
contributes to (\ref{p4}) to the total wave function with a contribution of
order $(\epsilon\,\hat{U}/V^2)\sqrt{1-V^2}$. According to the approximation
scheme defined in the beginning of the present section we have $V^2\gg
\epsilon\, U$. Since $\hat{U}$ and $U$ are of same order of magnitude, the
radiated part is, as expected, a small quantity.

\section{Analysis of the results}\label{analysis}

 In this section we analyse the solutions of 
Eqs (\ref{p7}),(\ref{s3.14}), and (\ref{s3.15}), 
(\ref{s3.13}),(\ref{s3.16bis}) and (\ref{s3.16.b}) which describe the
dynamics of the system within our approach. The separation between the slow
and fast times we used up to now in order to identify which time derivatives
were negligible is no longer necessary, and we will henceforth only employ the
actual time $t$. We will also drop the multiplicative factor $\epsilon$ in
front of the perturbing potential $U(z)$ and of $\phi_1(x,t)$.
In the two following subsections we study the evolution of the
parameters of the soliton and in the two last ones we analyse the radiated
part. 

\subsection{Effective potential approximation}\label{eff}
  
  Since we now use the actual time $t$, instead of using
Eqs.(\ref{p7}),(\ref{s3.14}) and (\ref{s3.15}), it is more appropriate to work with the equivalent equations (\ref{s2.6.1},\ref{s2.6.2}).
In order to get insight into the details of the dynamics of the soliton,
one should solve these equations numerically for a particular obstacle. This
is done in section \ref{num}, where we study the behavior of a soliton
incident on a delta scatterer. But before going to this point, it is
interesting to study some limiting cases. In particular, the dynamics of the
variationnal solution (\ref{p1a}) can be more easily understood in the limit
of a very dark soliton (almost back).
To this end, let us 
multiply Eq.~(\ref{s2.6.1}) by $\dot{\bar{z}}$ and
add it to Eq.~(\ref{s2.6.2}) multiplied by  $\dot\theta$. This gives
\begin{equation}\label{s2.7}
4\,\dot{\theta}\,\sin\theta\,\cos^2\theta  =  
2\;\mbox{Re}\;\left\{\int_{-\infty}^{+\infty}\!\!\!dz \;
R^*[\phi_{\mbox{\tiny{sol}}}]\;\left(
\dot{\bar{z}}\,\partial_{\bar{z}} \phi_{\mbox{\tiny{sol}}} +
\dot{\theta}\,\partial_\theta \phi_{\mbox{\tiny{sol}}}\right)
\right\} \; .
\end{equation}

  In the limit of a weak potential, we have to keep in mind that $R$ is a
small quantity (of order of $U$). It is then legitimate at first order to
replace on the RHS of (\ref{s2.7}) $\dot{\bar{z}}$ by $\sin\theta$ and
to drop the term $\dot\theta$. One thus obtains
\begin{equation}\label{s2.8}
\dot\theta=\frac{3}{4}\cos^2\theta\,\int_{-\infty}^{+\infty}\!\!\!dz\,
\frac{\partial_zf}{\cosh^4[\cos\theta(z-\bar{z})]} \; .
\end{equation}

A further simplification of the equations is obtained in the limit of very
dark soliton, when $\theta\to 0$. In this limit
$\dot \theta \simeq \ddot {\bar{z}}$ and using expression (\ref{s1.7}) for $f$ we
can put Eq.~(\ref{s2.8}) in the following form:
\begin{equation}\label{s2.9}
2\,\ddot{\bar{z}}=-\frac{d U_{\mbox{\tiny eff}}}{d \bar{z}}
\qquad\mbox{where}\qquad
U_{\mbox{\tiny eff}}(\bar{z})
=-\frac{3}{2}\int_{-\infty}^{+\infty}\!\!\!dz\,
\frac{\delta f(z)}{\cosh^4(z-\bar{z})}
=\frac{1}{2}\,\int_{-\infty}^{+\infty}\!\!\!dz\,
\frac{U(z)}{\cosh^2(z-\bar{z})} \; .
\end{equation}
If we furthermore consider a potential $U(z)$ which slowly depends on $z$
(over a length scale much larger than unity \cite{rem3}), then $U(z)$ in the
convolution of the RHS of Eq.(\ref{s2.9}) does not appreciably vary over the
distance where the term $\cosh^{-2}(z-\bar{z})$ is noticeable. This yields
\begin{equation}\label{s2.10}
U_{\mbox{\tiny eff}}(\bar{z})\simeq
\frac{U(\bar{z})}{2}\,\int_{-\infty}^{+\infty}\!\!\!dz\,
\frac{1}{\cosh^2(z-\bar{z})} =U(\bar{z})\; .
\end{equation}
Equations(\ref{s2.9}) and (\ref{s2.10}) show than in the appropriate limit (very dark
soliton, weak and slowly varying potential) the soliton can be considered as
an effective classical particle of mass $2$ (i.e., twice the mass of a bare
particle) of position $\bar{z}$ (the position of the center of the density
trough) evolving in a potential $U(\bar{z})$. If we relax the hypothesis of
slowly varying potential, the soliton can still be considered as a particle of
mass 2, but it now evolves in an effective potential $U_{\mbox{\tiny
eff}}(\bar{z})$ defined in Eq.(\ref{s2.9}) as a convolution of the real potential
$U(z)$. The fact that the effective mass of the soliton is twice the one of a
bare particle has already been obtained in
Refs.~\cite{Mur99,Bus01,Fra02,Kon04}. Previous studies mainly focused on
slowly varying external potentials and, as a result, the existence of an
effective potential $U_{\mbox{\tiny eff}}$ -- different from $U$ -- had not
been noticed so far, except in Ref.~\cite{Fra02} where this result has already
been obtained in the special case of a $\delta$ scatterer. In the
following, we denote the approximation corresponding to Eq.~(\ref{s2.9}) as
the effective potential approximation: the soliton is considered as an
effective classical particle of mass 2, position $\bar{z}$, moving in the
potential $U_{\mbox{\tiny eff}}(\bar{z})$.

\subsection{Numerical check}\label{num}

Let us now study in detail a particular example. We consider a soliton
incident on a pointlike obstacle-i.e., a $\delta$ scatterer characterized by
$U(x)=\lambda\,\delta(x)$. In this case, the static background $f(z)$ is given
by Eq(\ref{s1.5}) and Eqs. (\ref{s2.6.1}) and (\ref{s2.6.2}) read
\begin{eqnarray}\label{s2.12}
\dot{\theta}&=&\mbox{sgn}(\lambda)
\cos^2\theta\int_0^{+\infty}\!\!\!\frac{dz}{\sinh(2z+2a)}
\left(\frac{1}{\cosh^4X}-\frac{1}{\cosh^4Y}\right) \nonumber \\
& + &
\frac{\cos^3\theta}{2}\int_0^{+\infty}\!\!\!dz\,\big[1-f^2(z)\big]
\left(\frac{\tanh X}{\cosh^4X}-\frac{\tanh Y}{\cosh^4Y}\right) 
\; ,\end{eqnarray}
\noindent and
\begin{eqnarray}\label{s2.13}
\sin{\theta}-\dot{\bar{z}}
&=&\mbox{sgn}(\lambda)
\sin\theta\int_0^{+\infty}\!\!\!\frac{dz}{\sinh(2z+2a)}
\left(\frac{X\cosh^{-2}X+\tanh X}{\cosh^2X}
+\frac{Y\cosh^{-2}Y+\tanh Y}{\cosh^2Y}
\right) \nonumber \\
& + &
\frac{\sin\theta\cos\theta}{2}
\int_0^{+\infty}\!\!\!dz\,\big[1-f^2(z)\big]
\left(\frac{1-X\tanh X}{\cosh^4X}+\frac{1-Y\tanh Y}{\cosh^4Y}\right) 
\; .
\end{eqnarray}
$X$ and $Y$ in Eqs. (\ref{s2.12}) and (\ref{s2.13}) are notations for
$(z-\bar{z})\cos\theta$ and $(z+\bar{z})\cos\theta$ respectively, and the
expressions of function $f$ and of parameter $a$ are given in
Eq.~(\ref{s1.5}). Solving Eqs. (\ref{s2.12}) and (\ref{s2.13}) numerically, we
obtain the time evolution of the parameters of the soliton. We plot in Figs. 2
and 3 the behavior of $\bar{z}$ as a function of $t$ for different initial
velocities $V$. Figure 2 corresponds to a repulsive interaction with
$\lambda=+1$ and Fig. 3 to an attractive one with $\lambda=-1$. The initial
conditions for the numerical integration of Eqs. (\ref{s2.12}) and (\ref{s2.13}) are
taken to be $\bar{z}(t=0)=-10$ and $\dot{\bar{z}}(t=0)=\sin[\theta(t=0)]=V$.
Several curves are drawn, corresponding to several values of $V$. In the
repulsive case (Fig. 2), three initial velocities have been chosen: $V=0.9$,
0.707, and 0.4. The value $V=0.9$ corresponds to a fast soliton which is weakly
perturbed by the barrier, the value $V=0.4$ corresponds to a reflected soliton, and the
value $V=0.707$ is just below the value $V=\sqrt{\lambda/2}$ which, according
to the effective potential approximation (\ref{s2.9}), is the separatrix
between transmission and reflexion (corresponding to
$V^2=\mbox{max}\;\{U_{\mbox{\tiny eff}}(\bar{z})\} = \lambda/2$). In the
attractive case (Fig.~3) the curves are drawn in the cases $V=0.707$, 0.4 and
0.3. In both figures, the solid lines correspond to the exact numerical
solution of Eqs.(\ref{s2.12}) and (\ref{s2.13}) and the dashed lines to the result of
the effective potential approximation.

We first remark that the case of a $\delta$ scatterer we consider here is the
worst possible for the effective potential approximation and that this
approximation is certainly more at ease with smoother potentials. However, it
is interesting to note that the effective potential approximation, which could
be thought as oversimplified, is often very good. The worst agreement occurs
in the case of repulsive obstacle, near the separatrix (which is estimated by
the effective potential approximation to occur in the case of Fig. 2 at
$V=1/\sqrt{2}$). As we will see below (Fig. 4), the effective potential
approximation does not exactly predict the location of this separatrix
whereas, in this region, the trajectories are strongly affected by small
changes of the initial velocity $V$. This is the reason for the bad agreement
of the result of the approximate method with the ones given by the numerical
integration of Eqs. (\ref{s2.12}) and (\ref{s2.13}) for $V=0.707$. However, it is
surprising to note that the effective potential approximation is generically
valid, even in the case where the soliton is far from being very dark: even
the limit $V\to 1$ is very accurately described by this approximation on
Figs. 2 and 3.


\begin{center}
\begin{minipage}[t]{7.5 cm}
\includegraphics[width=6cm]{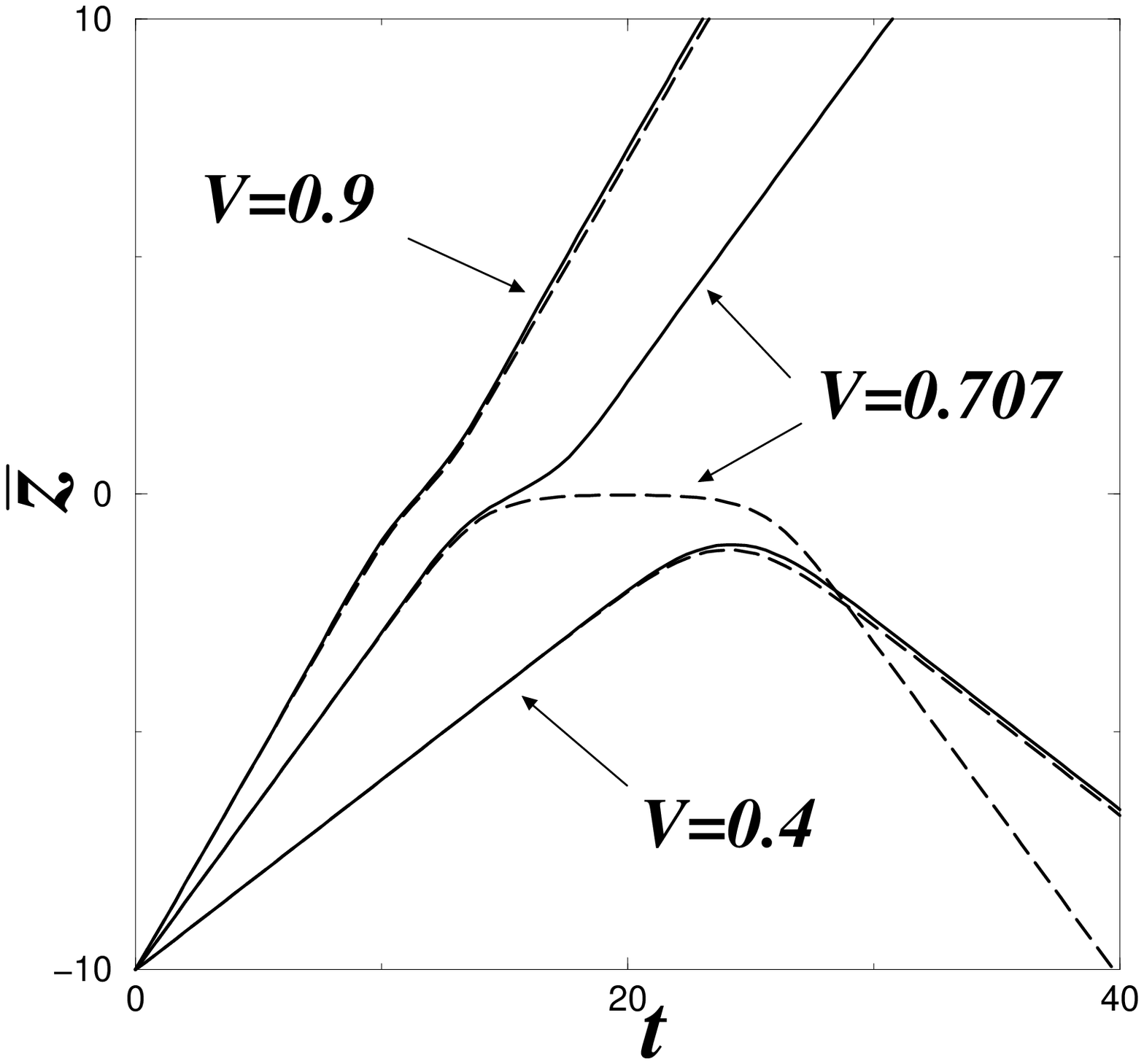}

FIG. 2. {\sl $\bar{z}(t)$ for solitons of initial velocity $V$, incident on a
repulsive obstacle $U(x)=\lambda\, \delta(x)$ with $\lambda=+1$. 
The solid lines correspond to the
numerical solution of Eqs.(\ref{s2.12}) and (\ref{s2.13}) and the dashed lines to
the effective potential approximation (\ref{s2.9}).}
\end{minipage}
\hspace{0.5 cm}
\begin{minipage}[t]{7.5 cm}
\includegraphics[width=6cm]{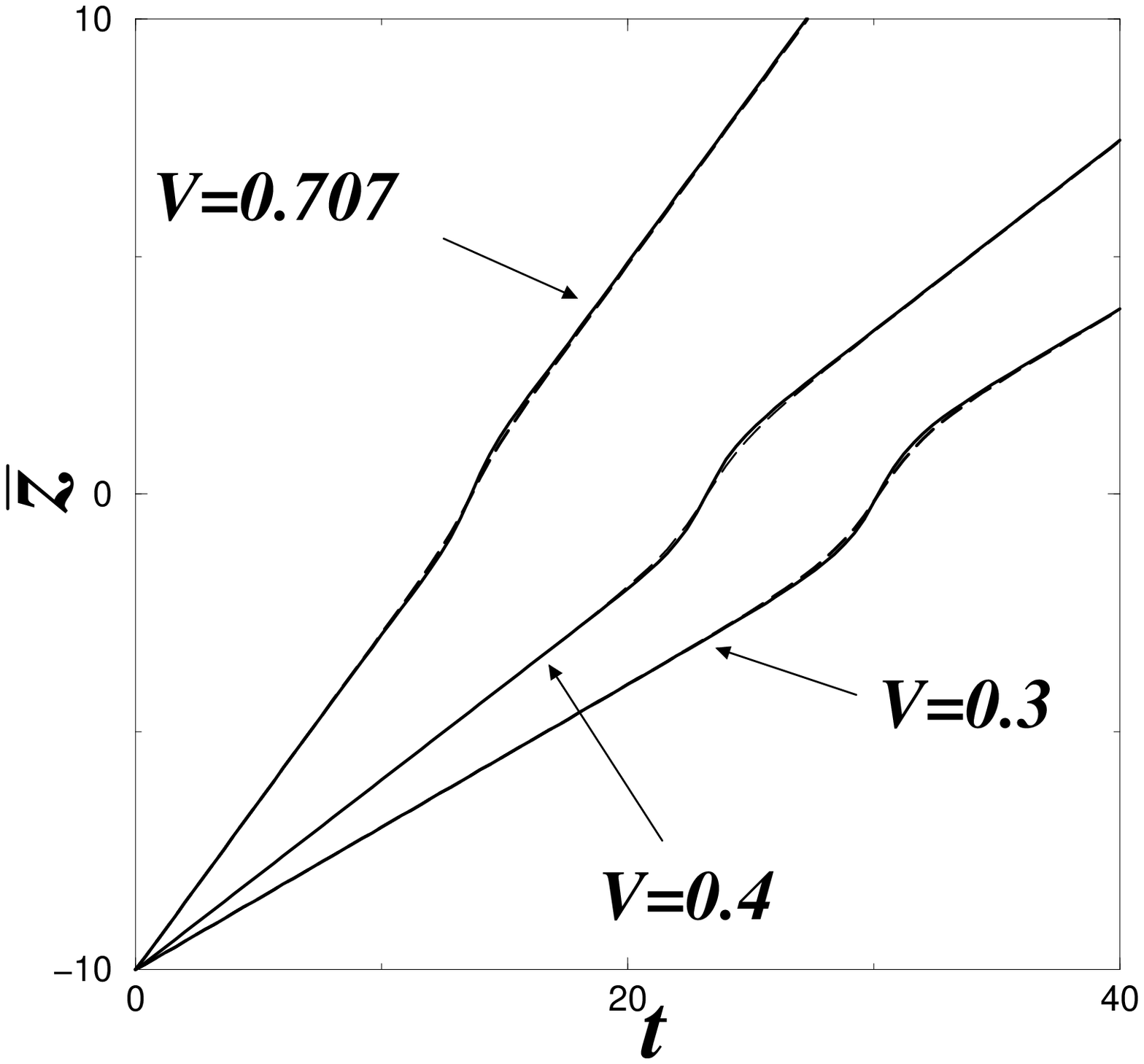}

FIG. 3.: {\sl Same as Fig. 2 for solitons in\-ci\-dent on an attractive
obstacle $U(x)=\lambda\,\delta(x)$ with $\lambda=-1$. The dashed lines
corres\-pon\-ding to the effective potential approximation are hardly
distinguishable from the solid lines which corres\-pond to the numerical
solution of Eqs.(\ref{s2.12}) and (\ref{s2.13}).}
\end{minipage}
\end{center}

In order to investigate more precisely the limit of large initial velocities
$V$ and to assess the validity of the effective potential approximation, let
us now establish the form of Eqs. (\ref{s2.6.1}) and (\ref{s2.6.2}) in the case of a
weakly perturbed soliton. From the effective potential approximation, one
infers that the soliton is weakly perturbed by the obstacle when its initial
energy is large compared to the external potential $U_{\mbox{\tiny eff}}$-
i.e., in the regime $V^2\gg U$ (since $U_{\mbox{\tiny eff}}$ and $U$ are
typically of same order of magnitude). This is confirmed by the
numerical results presented on Figs. 2 and 3: the trajectory of the soliton is
less modified for large $V$. In the extreme limit $V^2\gg U$ one may write
$\theta(t) = \theta_0 + \Theta(t)$ and $\bar{z}=V\,t + \Delta(t)$, with
$\Theta \ll \theta_0$ and $\dot{\Delta}\ll V$. $\Delta$ has the meaning of a
shift in position: it is the difference between the position of the center of
the soliton in presence of the obstacle with the value it would have in
absence of the obstacle. The perturbative versions of Eqs.
(\ref{s2.6.1}) and (\ref{s2.6.2}) read
\begin{equation}\label{s2.14}
4\,\dot{\Theta}\,\cos^2\theta_0 = 
-2\;\mbox{Re}\;\int_{-\infty}^{+\infty}\!\!\!dz \;
R^*[\Phi(z-Vt,\theta_0)]\;\Phi_z(z-Vt,\theta_0) \; 
\end{equation}
\noindent and
\begin{equation}\label{s2.15}
4\,\cos^2\theta_0\,[\Theta\cos\theta_0-\dot{\Delta}\, ]=
2\;\mbox{Re}\;\int_{-\infty}^{+\infty}\!\!\!dz \;
R^*[\Phi(z-Vt,\theta_0)]\;\Phi_\theta(z-Vt,\theta_0) \; .
\end{equation}
>From these equations it is a simple matter to compute analytically the
asymptotic expressions of the soliton parameter. One obtains -- as expected --
$\Theta(+\infty)=0$, and the asymptotic shift in position is
\begin{equation}\label{s2.16}
\Delta(+\infty)=
-\hat{U}(0) \, \frac{1+2\,V^2}{6\,V^2}\; ,
\end{equation}
\noindent where $\hat{U}(0)=\int_\RR dx\,U(x)$.
Equation (\ref{s2.16}) for $\Delta(+\infty)$ is an approximation (valid in
the regime $V^2\gg U$) of the exact result
\begin{equation}\label{s2.16.b}
\Delta(+\infty)=\lim_{t\to+\infty}\{\bar{z}(t)-Vt\}\; .
\end{equation}
Comparing definitions (\ref{p8}) and (\ref{s2.16.b}) we see that, since
$\sin\theta(t_1\to +\infty)=V$, one has  $\Delta(+\infty)=\tilde{z}(+\infty)$.
In the case of a $\delta$ scatterer, the exact value (\ref{s2.16.b}) was computed
through numerical solution of Eqs. (\ref{s2.12}) and (\ref{s2.13}). The result is
displayed in Fig.~4 (thick solid curves) and compared with the approximate
expression (\ref{s2.16}) (thin solid curves) and with the result of the
effective potential approximation (dashed curves).

\vspace{0.5 cm}

\begin{minipage}{8 cm}
\includegraphics*[width=6.5cm]{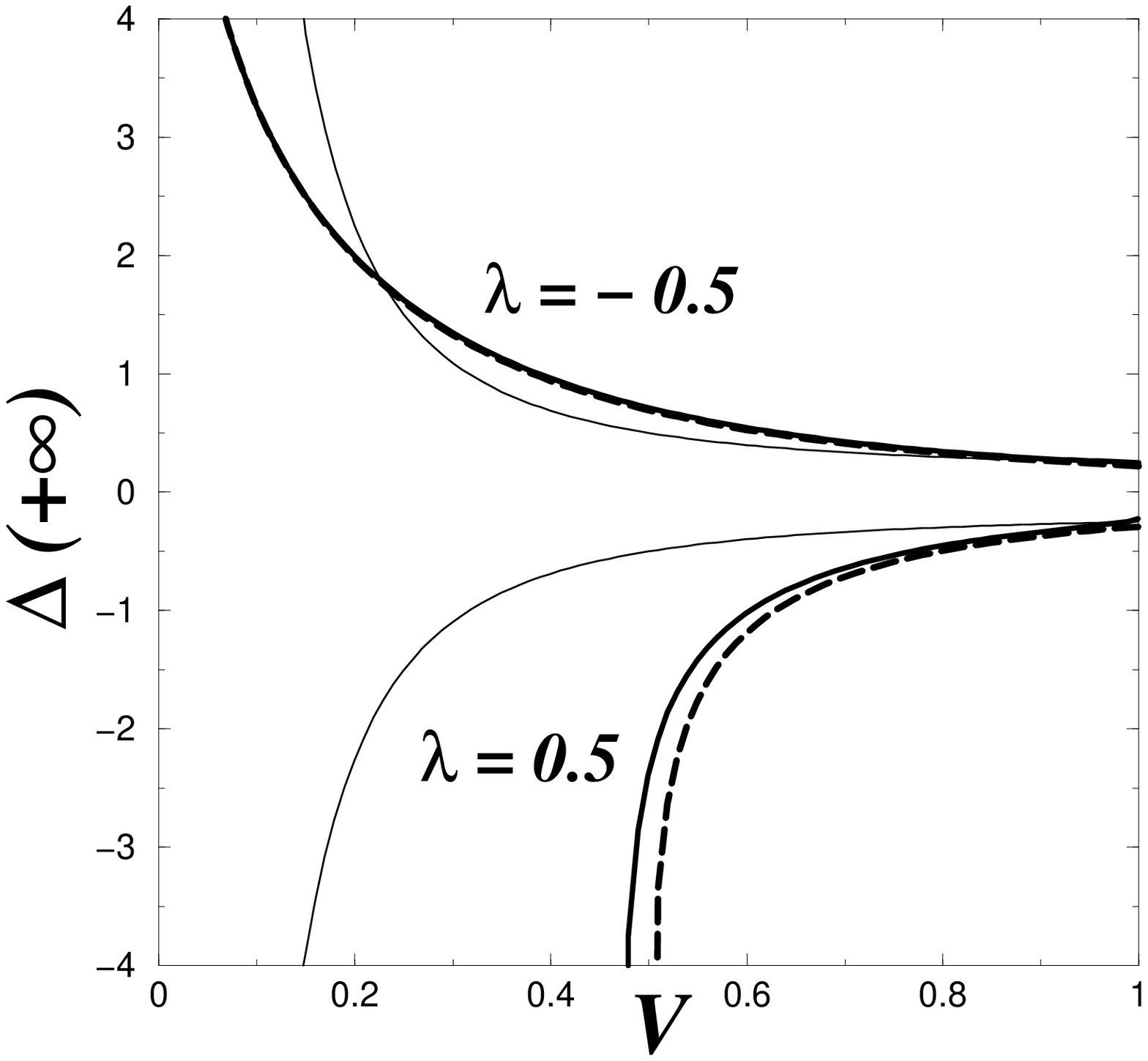}
\end{minipage}
\begin{minipage}{7.5 cm}
FIG. 4. {\sl $\Delta(+\infty)$ as a function of the initial velocity $V$
	of a soliton incident on a $\delta$ peak $U(x)=\lambda\,\delta(x)$. The upper
	curves correspond to the case $\lambda=-0.5$, the lower ones to the case
	$\lambda=0.5$. The thick solid lines are the exact result (\ref{s2.16.b})
	obtained form the numerical integration of Eqs. (\ref{s2.12}) and (\ref{s2.13}).
	The dashed curves are the result (\ref{s2.17}) of the effective potential
	approximation and the thin solid curves are the approximate result
	(\ref{s2.16}).}
\end{minipage}

\vspace{0.5 cm}

 In the case of the effective potential approximation, the value of the shift
$\Delta(+\infty)$ can be computed either via the numerical solution of the
equation of motion (\ref{s2.9}) or via the formula
\begin{equation}\label{s2.17}
\Delta(+\infty)=\int_{-\infty}^{+\infty}\!\!\!dx
\left[1-\frac{1}{\sqrt{1-U_{\mbox{\tiny eff}}(x)/V^2}} \right]\; .
\end{equation}
>From this expression, one sees that in the limit $V^2\gg U_{\mbox{\tiny eff}}
\sim U$, the effective potential approximation yields a result
$\Delta(+\infty)\simeq -\frac{1}{2}\,\hat{U}(0)/V^2$. Hence, in this limit,
the shift computed via the effective potential approximation at $V=1$ is
correct [since it agrees with the result (\ref{s2.16}) at $V=1$]. This is
surprising, because the effective potential approximation is expected to be
accurate only for very dark solitons. However, one can also notice that
detailed agreement with the exact result (\ref{s2.16.b}) is missed since, in
the limit $V^2\gg U$, the asymptotic evaluation (\ref{s2.16}) of
(\ref{s2.16.b}) does not exactly match the one of (\ref{s2.17}). Yet one sees
from Fig.~4 that the shift computed via the effective potential approximation
is in surprisingly good agreement with the exact value, even for fast
solitons. In particular, in the case of an attractive potential, the exact
evaluation of $\Delta(+\infty)$ and its approximation (\ref{s2.17}) are
hardly distinguishable.

\subsection{Backward- and forward- emitted wave packets}\label{bfwp}

  At this point it is interesting to study in more detail the structure of the
phonon part of the wave function-i.e., of $\phi_1(x,t)$.
>From Eqs.
(\ref{s3.13}) and (\ref{b3}) one can separate $\phi_1$ into two parts:
$\phi_1=\phi_1^++\phi_1^-$ with
\begin{equation}\label{sd1}
\phi_1^\alpha(x,t)=\int_\RR\!\!\!dq\,C_q^\alpha(t)\,u_q^\alpha(x)\; .
\end{equation}
>From the
explicit expressions (\ref{s3.16ter}), (\ref{s3.16bis}) and (\ref{b3}) one sees
that $\phi^+_1$ ($\phi^-_1$) describes waves propagating toward the positive
(negative) $x$. 

We are interested in studying the outcome of the collision-i.e., in obtaining
an analytical evaluation of Eq(\ref{sd1}) when $t\to +\infty$. To this end, one
uses the fact that at large time one has $C_q^\alpha(t) \propto \exp
\{-i\varepsilon_q^\alpha [t+\tilde{z}(+\infty)/V]\}$. Hence, instead of
working with the variable $t$, it is convenient here to define
$\tau=t+\tilde{z}(+\infty)/V$ and to write Eq.~(\ref{sd1}) in the
 form
\begin{equation}\label{sd1a}
\phi_1^\alpha(x,t)=\int_\RR\!\!\!dq\,G^\alpha(q,x)\,
\exp\{i[q(x+V\,\tau)-\alpha\,\tau\,F(q)]\}\; ,
\end{equation}
\noindent where $F(q)=q\,(q^2/4+1)^{1/2}$ and
$G^\alpha(q,x)=[q/2+\varepsilon_q^\alpha/q+i\chi(x)]^2
D_q^\alpha(+\infty) \exp
\{i\varepsilon_q^\alpha \tilde{z}(+\infty)/V\}$. In
the appropriate limit (to be defined soon), one can evaluate this expression
through a saddle phase estimate. In this limit, the rapidly oscillating 
phase in Eq.(\ref{sd1a}) is stationary at
point $\pm q_\alpha$ which are solutions of $x+V\,\tau=\alpha\,\tau\,F'(q)$. 
One has
\begin{equation}\label{sd2}
q_\alpha^2=\frac{1}{2}\left[ X^2-4+\alpha\,X\,\sqrt{X^2+8} \, \right] \; ,
\quad\mbox{with}\quad X=V+\frac{x}{\tau}\; .
\end{equation}
One can easily verify that $q_\alpha$ goes to zero when
$V+\frac{x}{\tau}=\alpha$ and that $q_\alpha^2$ is positive only if
$\alpha(V+\frac{x}{\tau})>1$. From this, one sees that the saddle phase
estimate of Eqs.(\ref{sd1}) and (\ref{sd1a}) is accurate when the two saddles are well
separated-i.e., in the regime $x\gg(1-V)\,\tau$ for $\alpha=+$ and
$x\ll-(1+V)\,\tau$ for $\alpha=-$. If this condition is fulfilled, one
obtains
\begin{eqnarray}\label{sd3}
\phi_1^\alpha(x,t)& \simeq &
G^\alpha(q_\alpha,x)\,\sqrt{\frac{2\,\pi}{|F''(q_\alpha)|}}
\,\ep^{i[q_\alpha(x+V\,\tau)-\alpha\,\tau\,F(q_\alpha)-\alpha\,\pi/4]}
\nonumber \\
& + & 
G^\alpha(-q_\alpha,x)\,\sqrt{\frac{2\,\pi}{|F''(q_\alpha)|}}
\,\ep^{-i[q_\alpha(x+V\,\tau)-\alpha\,\tau\,F(q_\alpha)-\alpha\,\pi/4]} \; .
\end{eqnarray}
The exact expression computed from Eq.(\ref{sd1}) is compared in Fig.~5 with the
saddle phase estimate (\ref{sd3}). The curves are drawn at $\tau=60$
\cite{temps} for a soliton with incident velocity $V=0.5$. The obstacle is
here taken to be a delta scatterer $\lambda\,\delta(x)$. 
$\phi_1$ being proportional to $\lambda$ [through the expression (\ref{s3.20}) of
$D_q^\alpha(+\infty)$] we represent in Fig.~5 the value of $\phi_1(x,t)/\lambda$
(actually its real part) which do no depend on $\lambda$.

\

 One sees in Figure 5 that the semiclassical approximation (\ref{sd3}) is
excellent in all its expected domain of validity and diverges at
$x=(1-V)\,\tau=30$ (for $\alpha=+$) and $x=-(1+V)\,\tau=-90$ (for $\alpha=-$)
\cite{coal}. Hence, these points can be considered as representative of the
region where the contribution of $\phi^+_1$ and $\phi_1^-$ to the total wave
function is more important. Roughly speaking, the present approach indicates
that, long after the collision, $\phi_1^\alpha(x,t)$ is maximum around
$x=(\alpha-V)\,\tau$. We recall that when using $x$ (instead of $z$) as
position coordinate, the soliton is, at all times, located around $x=0$.
Hence, going back to the $z$ coordinate, we have a clear picture of the
process at large times: the soliton propagates at velocity $V$ (the same as
its initial velocity) after having emitted phonons which form two wave
packets, one propagating in the forward direction with group velocity 1 (i.e.,
the sound velocity) and the other one propagating backwards with group
velocity $-1$. The same conclusion seems to be reached in the numerical
simulations of Parker {\it et al.} \cite{Par03,Par04}.

\noindent
\begin{minipage}{9.7 cm}
\includegraphics*[width=9cm]{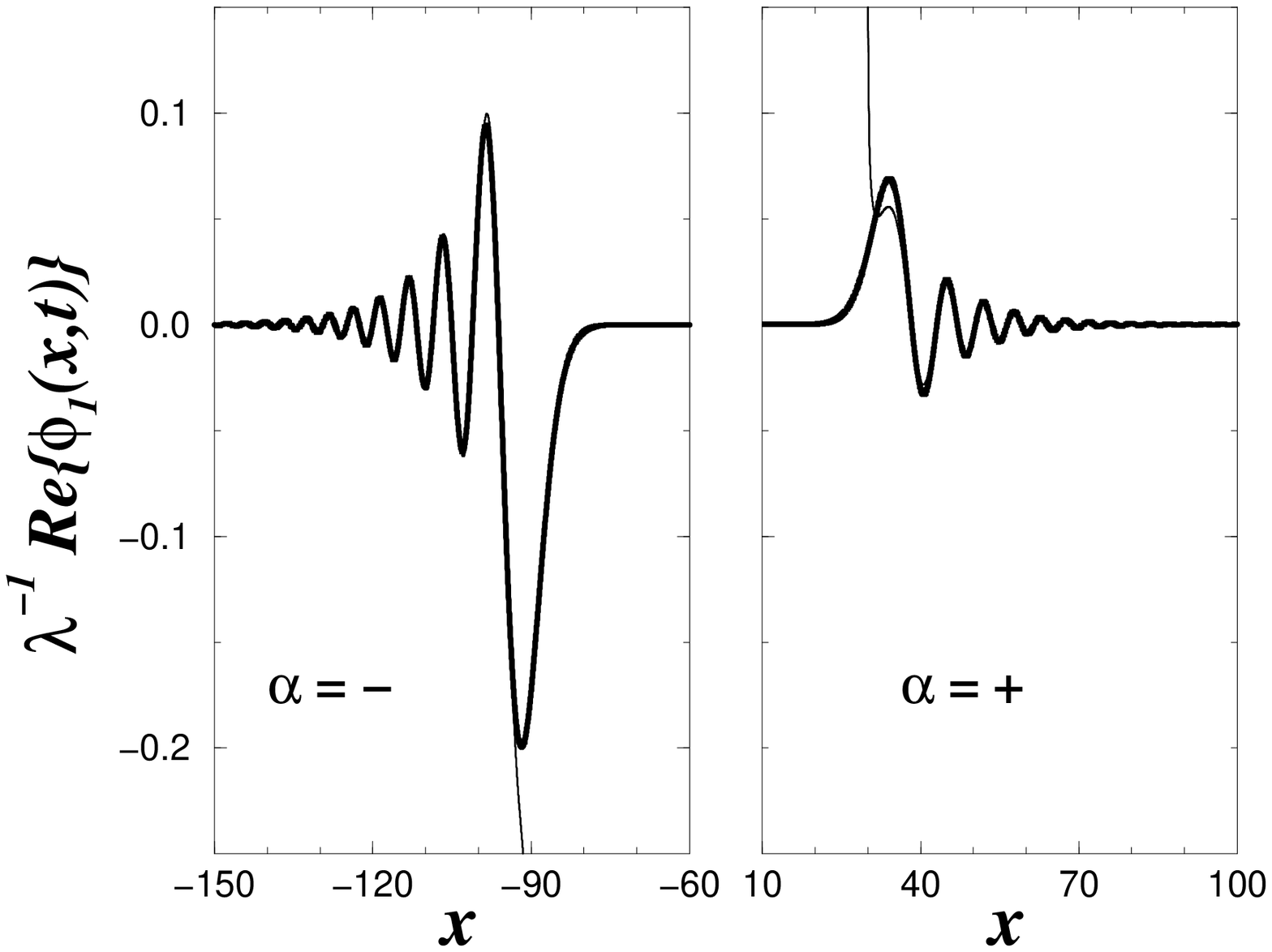}
\end{minipage}
\begin{minipage}{6.7 cm}
Fig. 5. {\sl $\mbox{Re}\,\{\phi_1(x,t)\}$ as a function of $x$ for $\tau=60$
	for a soliton of initial ve\-lo\-ci\-ty $V=0.5$ incident on a $\delta$
	scatterer $\lambda\,\delta(x)$.
	The thick line represents the result (\ref{sd1}) and the thin line its
	semiclassical approximation (\ref{sd3}). For le\-gi\-bi\-li\-ty we have
	se\-pa\-ra\-ted the region where $\phi_1^-$ is nonzero (around $x=-90$)
	from the one where $\phi_1^+$ is non zero (around $x=30$). Note that in the
	expected domain of validity of Eq.~(\ref{sd3}) ($x\gg 30$ and $x\ll -90$) one
	can hardly distinguish the thick line from its semiclassical approximation.}
\end{minipage}

\subsection{Radiated energy}\label{rad_ener}

A quantity of importance for characterizing the system is the total energy
radiated by the soliton. Equation(\ref{s1.3}) for the field $\psi$ which, in the
present work, is of the form $\varphi(z,t)\exp(-i\,t)$ [cf. Eq. (\ref{s1.8})],
conserves the energy ${\cal E}$ defined as:
\begin{equation}\label{e1}
{\cal E}[\varphi]=\int_\RR\!\!\!dz\,
\left\{\frac{1}{2}\,|\varphi_z|^2+\frac{1}{2}\,(|\varphi|^2-1)^2+
U(z)\,|\varphi|^2\right\}\; .
\end{equation}
In order to have an expression of the energy in terms of the field $\phi$
which, when $\phi=\Phi$, matches the usual expression (\ref{a5b}) of the energy
of the soliton, we rather work with the quantity $E[\phi]={\cal E}[f\phi]-
{\cal E}[f]$. $E[\phi]$ is of course a conserved quantity, and we are
interested in its expression far before ($t\to-\infty$) and far after
($t\to+\infty$) the collision with the obstacle. We note here that
$f(z)-1$ and $U(z)$ are non
zero only when $z$ is close to to origin, whereas, in the same region,
$\phi(z,t)-1$ is zero when $t\to\pm\infty$. After a change of variable from
$z$ to $x=z-\bar{z}(t)$, the previous remark allows one to obtain the
simplified expression for $E$ (only valid when $t\to\pm\infty$):
\begin{equation}\label{e2}
E[\phi]=\frac{1}{2}\,\int_\RR\!\!\!dx\,
\left\{|\phi_x|^2+ (|\phi|^2-1)^2\right\}\; .
\end{equation}
Using the decomposition (\ref{p1}), keeping the lowest orders in $\delta\phi$,
 and taking into account the fact that, when
$t\to\pm\infty$, $|\Phi|^2-1$ is zero in the regions where $\delta\phi$ is
noticeable, one obtains
\begin{equation}\label{e3}
E [\phi]=\frac{1}{2}\,\int_\RR\!\!\!dx\,
\left\{|\Phi_x|^2+ (|\Phi|^2-1)^2\right\}
+\frac{1}{2}\,\int_\RR\!\!\!dx\,
\left\{|\delta\phi_x|^2+\big(\Phi^*\delta\phi+\delta\phi^*\Phi\big)^2\right\}
+{\cal O}(\delta\phi^3) \; .
\end{equation}
The first integral on the RHS of Eq.(\ref{e3}) corresponds to the soliton's
energy and is equal to $\frac{4}{3}
\cos^3\theta$. The second integral on the RHS of Eq.(\ref{e3}) corresponds
to the energy of the radiated part and is denoted by $E_{\mbox{\tiny{rad}}}$
in the following. 

  We are now facing a difficulty: we performed a computation at order
$\epsilon$ and at this order we have $\theta(+\infty)=\theta(-\infty)$
since the equations for the parameters of the soliton are the same as the one
obtained in the adiabatic approximation (see the discussion at the end of
\ref{lagrangien}). Accordingly, $E_{\mbox{\tiny{rad}}}$
in Eq.(\ref{e3}) being of order $\epsilon^2$ should be neglected. Hence, at order
$\epsilon$ nothing has occurred for the soliton's energy: this quantity is not
modified by the collision with the obstacle and the radiated energy should be
neglected. Thus, it seems that our first-order approach is unable to predict
the amount of energy lost by the soliton during the collision with the
obstacle.

However, as already remarked in the study of the scattering of bright solitons
\cite{Kiv87}, one can circumvent this difficulty and extract some second-order
information from our results. The procedure is the following: when pushing the
computations at order $\epsilon^2$, the ${\cal O}(\epsilon^2)$ estimate of
$E_{\mbox{\tiny{rad}}}$ is still given by the second term on the RHS of
(\ref{e3}) with $\delta\phi=\epsilon\phi_1$, which we know from our first-
order approach. At second order, since $E_{\mbox{\tiny{rad}}}$ is non zero,
the soliton's energy has been modified by the collision and energy
conservation now reads
\begin{equation}\label{e4}
\frac{4}{3}\,\cos^3[\theta(-\infty)]=E=
\frac{4}{3}\,\cos^3[\theta(+\infty)]+E_{\mbox{\tiny{rad}}}\; .
\end{equation}
Equation (\ref{e4}) allows us to determine the change in the soliton's parameter
$\theta$. Writing $\theta(-\infty)=\theta_0$ (with $\sin\theta_0=V$) and
$\theta(+\infty)=\theta_0+\delta\theta$ one obtains
\begin{equation}\label{e5}
\delta\theta=\frac{E_{\mbox{\tiny{rad}}}}{4\,\cos^2\theta_0\sin\theta_0}
\; .
\end{equation}
 From Eq.~(\ref{e5}) one can also determine the velocity at $t\to +\infty$
 which is equal to $\sin[\theta(+\infty)] = V + \delta\theta \cos\theta_0$.
 Thus, we can determine how the collision has affected the soliton's shape and
 velocity by computing $E_{\mbox{\tiny{rad}}}$ (replacing $\delta\phi$ by
 $\phi_1$). This will be done in the remaining of this section.

   On the basis of the analysis in terms of forward- and backward-emitted wave
packet made in Section \ref{bfwp}, one can separate $E_{\mbox{\tiny{rad}}}$
into two parts, which we denote $E_{\mbox{\tiny{rad}}}^-$ and
$E_{\mbox{\tiny{rad}}}^+$, the first one corresponding to energy radiated
backwards and the second one to forward-radiated energy, with
\begin{equation}\label{e6}
E_{\mbox{\tiny{rad}}}^\alpha=\lim_{t\to+\infty}
\frac{1}{2}\,\int_\RR\!\!\!dx\,
\left\{|\delta\phi_x^\alpha|^2+\big(\Phi^*\delta\phi^\alpha
+\delta\phi^{\alpha\, *}\Phi\big)^2\right\}
\; .
\end{equation}
A long computation which is summarized in \ref{calcul-ener-rad} yields the
result
\begin{equation}\label{e7}
E_{\mbox{\tiny{rad}}}^\alpha=16\,\pi\,\int_0^{+\infty}\!\!\!dq\;
|D_q^\alpha(+\infty)|^2\,(\varepsilon_q^\alpha)^2\,
\left(\frac{q^2}{4}+1\right)
\; .
\end{equation}
When $D_q^\alpha(+\infty)$ is given by Eq.(\ref{s3.20}), one obtains
\begin{equation}\label{e8}
E_{\mbox{\tiny{rad}}}^\alpha=\frac{\pi}{16\,V^6}\int_0^{+\infty}\!\!\!dq\;
\frac{q^2 \, (\varepsilon_q^\alpha)^2
|\hat{U}(\varepsilon_q^\alpha/V)|^2}
{\sinh^2
\left(\frac{\ds\pi\,q\,\sqrt{1+q^2/4}}{\ds 2\,V\,\sqrt{1-V^2}}\right)
}
\; .
\end{equation}
The behavior at low and high velocity of
$E_{\mbox{\tiny{rad}}}^\alpha$ defined in Eq.(\ref{e8}) is the following
\begin{equation}\label{e9}
E_{\mbox{\tiny{rad}}}^\alpha \sim
\frac{\pi}{16\,V}\,\int_0^{+\infty}\!\!
\frac{q^4\,|\hat{U}(q)|^2}{\sinh^2\left(\frac{\pi\,q}{2}\right)}\,d q
\quad\mbox{when}\quad V\to 0\; ,
\end{equation}
\noindent and
\begin{equation}\label{e10}
E_{\mbox{\tiny{rad}}}^-\sim \frac{4}{15}\,(1-V^2)^{5/2}\,
|\hat{U}(0)|^2 
\quad , \quad
E_{\mbox{\tiny{rad}}}^+\sim \frac{2}{35}\,(1-V^2)^{9/2}\,
|\hat{U}(0)|^2 \; ,
\quad\mbox{when}\quad V\to 1\; .
\end{equation}
One sees from Eq.~(\ref{e9}) that our approach predicts an unphysical
divergence of the radiated energy at low incident soliton velocity. On the
contrary, numerical computations indicate that a soliton with very low
velocity does not radiate \cite{Par03,Par04}. However, one must bear in mind
that (\ref{e8}) is the result of a first-order expansion only valid in the
limit $V^2\gg U$ and is unable to tackle the regime of very low
incident velocities. More interestingly, in the high-velocity regime -- where
the first-order perturbation theory is valid -- we see from Eq.~(\ref{e10})
that the leading-order estimate of the total amount of radiation (forward or
backward emitted) vanishes.

In order to fix the ideas, we plot in Fig. 6 the value of
$E_{\mbox{\tiny{rad}}}^\alpha$ as a function of the initial soliton velocity
$V$. The obstacle is here taken to be a$\ delta$ scatterer $U(z) =
\lambda\,\delta(z)$. In this case $\hat{U}(q)=\lambda$.

\vspace{0.7 cm}

\begin{minipage}{8 cm}
\includegraphics*[width=7cm]{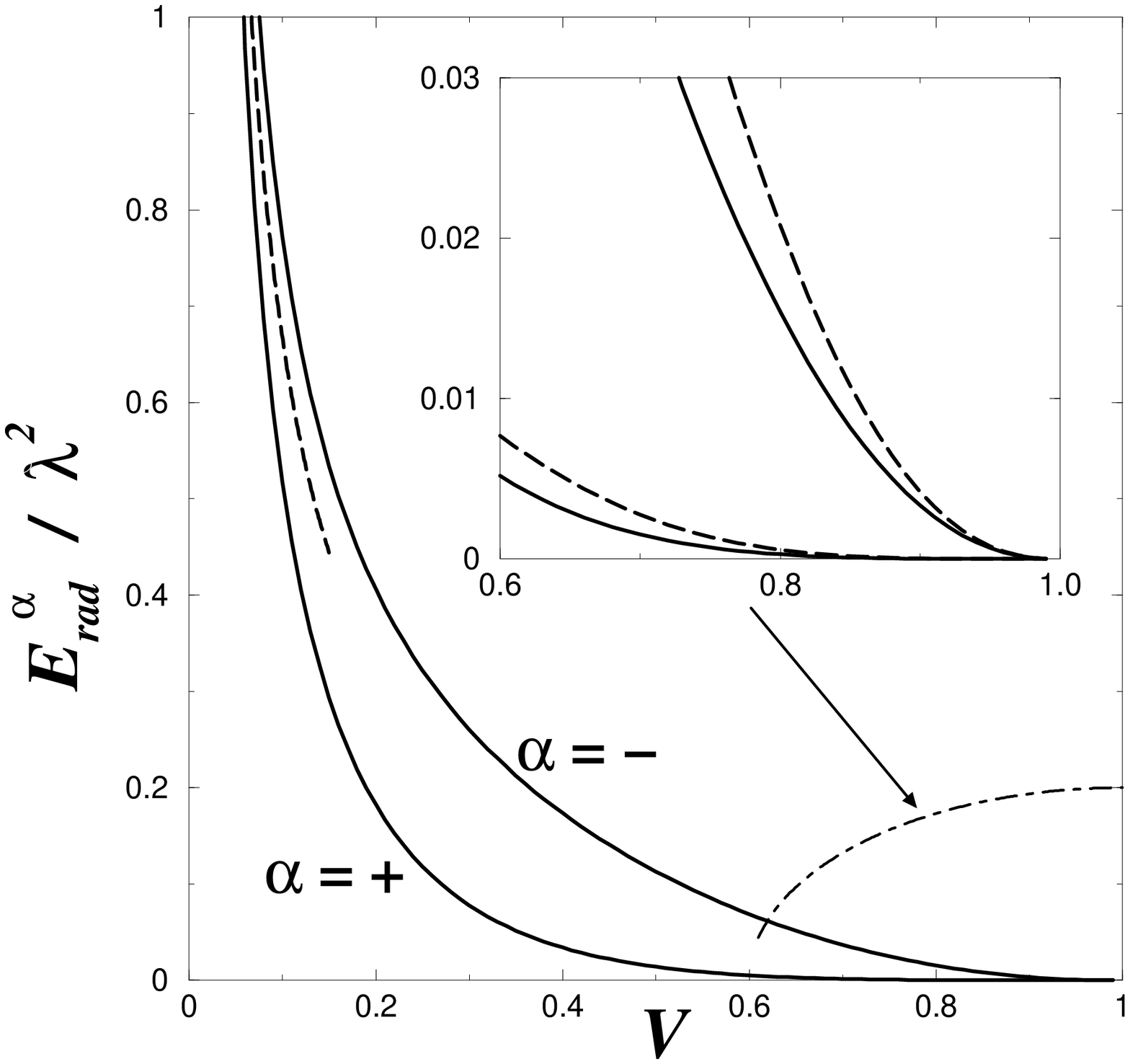}
\end{minipage}
\begin{minipage}{6.5 cm}
FIG 6. {\sl Energy $E_{\mbox{\tiny{rad}}}^\alpha$ radiated in the forward
	($\alpha=+$) and backward ($\alpha=-$) directions by a soliton of
	initial ve\-lo\-ci\-ty $V$ incident on a $\delta$ scatterer. The solid lines
	represent the result (\ref{e8}) and the dashed line the approximation
	(\ref{e9}) which reads here $E_{\mbox{\tiny{rad}}}^\alpha \simeq
	\lambda^2/(15\,V)$. The inset displays a blowup of the fi\-gu\-re at high
	velocity. In the inset, the dashed curves are the asymptotic results
	(\ref{e10}).}
\end{minipage}

Figure 6 shows that most of the energy is radiated backwards (this was already
implicit in Fig. 5) and confirms that, at leading order in $U/V^2$,
a soliton does not radiate in the limit $V\to 1$. Besides, not only the
absolute value of $E_{\mbox{\tiny{rad}}}$ goes to zero, but also the relative
amount of energy radiated $E_{\mbox{\tiny{rad}}}/E$ vanishes [as $(1-V^2)$].
Very similar results are obtained for an obstacle interacting with the beam
through a finite-range potential (for instance, a Gaussian). This absence of
radiation of a fast soliton can be explained intuitively as follows: whatever
the sign of the potential describing the obstacle, the soliton loses energy
under the form of radiated phonons. Accordingly it gets less dark
[$\delta\theta>0$ in Eq. (\ref{e5})] and is accelerated. This increased
velocity after a loss of energy is a typical feature of dark solitons which
are sometimes referred to as effective particles having a negative kinetic
mass which decreases with increasing energy \cite{Fed99}. However, our results
show that, since the soliton velocity cannot exceed the speed of sound, a
soliton whose velocity is close to this upper limit cannot be further
accelerated and the radiative process is suppressed.

\section{Conclusion}\label{conclusion}

  In this paper we have presented a study of the dynamics of a dark soliton
experiencing a collision with a finite-size potential in a quasi-1D
condensate. We determined the evolution of the soliton's parameters
and also included radiative effects within secular perturbation theory.

  A first output of the present work is what we called the `` effective
potential theory'': in many instances the soliton can be described as an
effective classical particle of mass 2 (twice the mass of a bare particle)
evolving in an effective potential $U_{\mbox{\tiny eff}}$ [defined in
Eq.~(\ref{s2.9})]. This approximation is rigorously valid in the case of a
slow soliton incident on a weak potential, but its actual regime of validity
appears to be quite broad.

  The effective potential theory is an approximation where -- as in all
adiabatic approaches -- radiative effects are neglected. Perturbation theory
allows one to get a deeper insight into the collisional process and to
determine the amount of radiated energy at leading order in $U/V^2$.
We show that the radiated waves form two counterpropagating phonon wave
packets, and we predict that the radiative process is suppressed in the limit
of a soliton moving with a velocity close to the velocity of sound. This
result should be checked numerically; work in this direction is in progress.

Whereas adiabatic theory predicts that the soliton's shape and velocity are the
same far before and far after the collision with the obstacle, it is an
important feature of the perturbative approach of being able to determine
finite asymptotic modifications of the soliton's parameters due to the
collision. We computed [in Eq.~(\ref{e5})] the modification of the soliton's
parameters at leading order in $U/V^2$. The qualitative picture of
the collisional process drawn from our approach is the following: the soliton
radiates energy, gets less dark, and is accelerated. Since the velocity of a
dark soliton cannot exceed the velocity of sound in the system, it is natural
that this velocity appears as a threshold for emission of radiations. Roughly
speaking, a soliton with a velocity close to the velocity of sound cannot
radiate [as seen from Eqs.~(\ref{e10})] since its velocity cannot further
increase.

\bigskip

\noindent {\large \bf Acknowledgments}

\bigskip \noindent It is a pleasure to thank E. Bogomolny, C. Schmit,
G. Shlyapnikov, and C. Texier for fruitful discussions. We acknowledge support from CNRS and Minist\`ere de la Recherche (Grant ACI Nanoscience 201). Laboratoire de Physique Th\'eorique et Mod\`eles Statistiques is Unit\'e Mixte de Recherche de
l'Universit\'e Paris XI et du CNRS, UMR 8626.

\appendix
\section{}\label{modespropres}
\setcounter{equation}{0} 
In this appendix we present the eigenvectors and eigenvalues of the
Hamiltonian ${\cal H}$ defined in Eq.(\ref{s3.10}). ${\cal H}$ is
not diagonalizable because, as we will see below, 
its null space and the one of ${\cal H}^2$ are not
identical. If we denote by its ``generalized null space'' \cite{Wei85} the
union of these two null spaces, one can easily verify that it is spanned by the
four vectors $|\omega_o\rangle$, $|\omega_e\rangle$, $|\Omega_o\rangle$, and 
$|\Omega_e\rangle$ defined as
\begin{eqnarray}\label{b1}
|\omega_o\rangle=\left(
\begin{array}{r}
\Phi=\;\;\;\chi+i\,\sin\theta\\
-\Phi^*=-\chi+i\,\sin\theta
\end{array}\right) \; 
&\!\!\!\! ,\!\! &
|\Omega_o\rangle=\left(
\begin{array}{c}
x\,\chi_x+\chi\\
x\,\chi_x+\chi
\end{array}\right) \; 
, \nonumber \\
|\omega_e\rangle=\left(
\begin{array}{c}
\Phi_x=\chi_x\\
\Phi_x=\chi_x
\end{array}\right) \; 
&\!\!\!\! ,\!\! &
|\Omega_e\rangle=\left(
\begin{array}{l}
i\cos\theta\,\Phi_\theta=
-\cos^2\theta-i\sin\theta\,(x\,\chi_x+\chi)\\
i\cos\theta\,\Phi^*_\theta=
\;\;\;\cos^2\theta-i\sin\theta\,(x\,\chi_x+\chi)
\end{array}\right) \; ,
\end{eqnarray}
\noindent where the function $\chi(x,\theta)$ is defined in Eq.(\ref{s1.12}).
The kets defined in Eq.(\ref{b1}) verify ${\cal H}|\omega_o\rangle ={\cal
H}|\omega_e\rangle=0$ and ${\cal H}^2|\Omega_o\rangle={\cal
H}^2|\Omega_e\rangle=0$, with ${\cal H}| \Omega_o
\rangle=2\cos^2\theta\,|\omega_o\rangle$ and ${\cal
H}|\Omega_e\rangle=\cos^2\theta\,|\omega_e\rangle$. One sees from Eq.(\ref{b1})
that $|\omega_e\rangle$ and $|\Omega_e\rangle$ are, respectively, linked to
variations of the center of the soliton and of the parameter $\theta$
(i.e., to the phase change across the soliton): this is the reason why the
terms in $\theta_{t_1}$ and $\bar{z}_{t_1}$ in Eq.~(\ref{p9}) can be
rewritten in Eq.~(\ref{p10}) by means of $|\omega_e\rangle$ and
$|\Omega_e\rangle$. One can similarly show that $|\omega_o\rangle$ is linked
to modulations of the global phase of the soliton and that $|\Omega_o\rangle$
is linked to variations of the background density at infinity.

The remainder of the spectrum of ${\cal H}$ is what we call the ``phonon
spectrum.'' It has two branches which we denote ``$+$'' and ``$-$.''
The corresponding eigenvectors and eigenvalues are denoted $|\Xi^{\pm}_q\rangle$
and $\varepsilon_q^\pm$ with 
\begin{equation}\label{bb}
{\cal H}\,|\Xi^{\pm}_q\rangle=
\varepsilon_q^\pm\,|\Xi^{\pm}_q\rangle \; .
\end{equation} The explicit expression of the eigenvalues is given in the
main text [Eq. (\ref{s3.16ter})]. It can be simply obtained by considering the
form of Eq.~(\ref{bb}) when $x\to\pm\infty$. In this limit, $\Phi$ goes to a
constant, and looking for the eigenvectors under the form of plane waves,
$\exp\{i\,q\,x\}\, (U_q^\pm,V_q^\pm)^{\sss T}$ (where $U_q^\pm$ and $V_q^\pm$
are constants), yields the result (\ref{s3.16ter}). This is the reason why we
denote these excitations as phonons. A better denomination should be
``Bogoliubov excitations'' because, far from the soliton, 
their form and dispersion relation correspond indeed to the elementary
excitations of a constant background moving with velocity $-V$.

The exact expression (valid for all $x\in\RR$) of the eigenvectors is given by
the squared Jost solutions of the inverse problem \cite{Chen98}. They read
$|\Xi^{\pm}_q\rangle=(u_q^\pm(x),v_q^\pm(x))^{\sss T}$ with
\begin{eqnarray}\label{b3}
u_q^\pm(x) & = &
\exp\{i\, q\, x\} \left(\frac{q}{2}+\frac{\varepsilon_q^\pm}{q}
+i\,\chi\right)^2 \; ,\nonumber \\
v_q^\pm(x) & = &
\exp\{i\, q\, x\} \left(\frac{q}{2}-\frac{\varepsilon_q^\pm}{q}
+i\,\chi\right)^2 \; .
\end{eqnarray}
The natural inner product of two kets is $\langle \,\cdot\,
|\sigma_3|\,\cdot\,\rangle$, where $\sigma_3$ is the third Pauli matrice.
The eigenvectors have the following normalization:
\begin{equation}\label{b4}
\langle\Xi_p^\beta|\sigma_3|\Xi_q^\alpha\rangle 
=\int_\RR d x \, \big[ u_p^{\beta\,*}(x)u_q^\alpha(x)
-v_p^{\beta\,*}(x)v_q^{\alpha}(x) \big]
= {\cal N}_q^\alpha\,
\delta_{\alpha,\beta}\,\delta(p-q)\; ,
\end{equation}
\noindent with
\begin{equation}\label{b5}
{\cal N}_q^\alpha=
16\,\alpha\,\pi\,q\,\sqrt{\frac{q^2}{4}+1}\,
\left(\frac{\varepsilon_q^\alpha}{q}\right)^2\; .
\end{equation}
In the main text we also use the following orthogonality
relations:
\begin{equation}\label{b6}
\langle\omega_e|\sigma_3|\omega_e\rangle=
\langle\Omega_e|\sigma_3|\Omega_e\rangle= 
\langle\omega_e|\sigma_3|\Xi^\alpha_q\rangle=
\langle\Omega_e|\sigma_3|\Xi^\alpha_q\rangle= 0 \; ,
\quad
\langle\omega_e|\sigma_3|\Omega_e\rangle= -4\cos^3\theta\; 
\end{equation}
\noindent and
\begin{equation}\label{b7}
\langle\omega_o|\sigma_3|\omega_o\rangle=
\langle\omega_o|\sigma_3|\omega_e\rangle=
\langle \omega_o|\sigma_3|\Xi^\alpha_q\rangle= 0 \; ,
\quad
\langle\omega_o|\sigma_3|\Omega_e\rangle=2\,i\sin\theta\cos\theta\; .
\end{equation}

\section{}\label{lagrangien}
\setcounter{equation}{0}

   In this appendix we briefly present the Lagrangian approach for dark
soliton of Kivshar and Kr\'olikowski \cite{Kiv95} and derive the
Lagrange equations (\ref{s2.6.1}) and (\ref{s2.6.2}). 

In absence of the perturbation $R[\phi]$, Eq. (\ref{s1.9}) can be derived
from the following Lagrangian density:
\begin{equation}\label{a1}
{\cal L}[\phi,\phi^*]=\frac{i}{2}\,
(\phi^*\phi_t-\phi\,\phi_t^*)(1-\frac{1}{|\phi|^2})
-\frac{1}{2}|\phi_z|^2-\frac{1}{2}(|\phi|^2-1)^2\; .
\end{equation}
Accordingly, the energy and momentum are defined by
\begin{eqnarray}\label{a2}
E&=&\int_{-\infty}^{+\infty}\!\!\! d z \;
\left[\phi_t\frac{\partial \cal L}{\partial \phi_t}+
\phi_t^*\frac{\partial\cal L}{\partial \phi_t^*}- \cal L\right]
=
\frac{1}{2}\int_{-\infty}^{+\infty}\!\!\! d z\;
\big[|\phi_z|^2+(|\phi|^2-1)^2\big] 
\; ,\nonumber \\
P&=&\int_{-\infty}^{+\infty}\!\!\! d z \;
\left[\phi_z\frac{\partial \cal L}{\partial \phi_t}+
\phi_z^*\frac{\partial\cal L}{\partial \phi_t^*}\right]
=\frac{i}{2}\int_{-\infty}^{+\infty}\!\!\! d z\;
(\phi\,\phi_z^*-\phi^*\phi_z)(1-\frac{1}{|\phi^2|})
\; .
\end{eqnarray}  
The Lagrangian density (\ref{a1}) is not {\it a priori} the most natural one leading to
Eq.~(\ref{s1.9}), but for the asymptotic boundary condition we are working
with ($|\phi|\to 1$ when $z\to \pm\infty$), it yields a finite value of the
energy and, besides, the energy and
momentum are now, for a field of the form $\phi(x-V\,t)$ (in
particular, in the case of a soliton), related by the relation $\delta E=
V\,\delta P$, indicating that the background contribution has been removed
and allowing one to treat the soliton as a classical particle-like object
\cite{Bar93,Kiv98}. For completeness, we note that, for a soliton, $\phi$ is
given by Eq.(\ref{s1.11}) and its energy and momentum defined in Eq.(\ref{a2}) have
the following expressions:
\begin{equation}\label{a5b}
E= \frac{4}{3} \,\cos^3\theta
\; \; ,\quad
P= \pi-2\,\theta-\sin(2\,\theta)
\; .
\end{equation}

Following Kivshar and Kr\'olikowski \cite{Kiv95}, one can obtain adiabatic
equations of motion for the soliton's parameters in the following way. Let us
consider a variational approximation of the type of Eq. (\ref{p1a}); the field
of the soliton is pa\-ra\-me\-trized with time dependent quantities $q_1(t)$,
..., $q_n(t)$ and has no other time dependence:  $\phi_{\mbox{\tiny{sol}}}
(z,t) = \phi(z , q_1(t),
..., q_n(t))$. One first defines the Lagrangian for the $q_i$'s as being
\begin{equation}\label{a6}
L(q_1,\dot{q}_1,...,q_n,\dot{q}_n)
=\int_{-\infty}^{+\infty}\!\!\!dz\;
{\cal L}[\phi_{\mbox{\tiny{sol}}},\phi^{*}_{\mbox{\tiny{sol}}}] \; .
\end{equation}
Then the quantities $\partial_{q_i}L$ and $\partial_{\dot{q}_i}L$ are computed
via
\begin{equation}\label{a7}
\partial_{q_i}L
=\int_{-\infty}^{+\infty}\!\!\! d z \; 
\big(\partial_{q_i}\phi\,\partial_\phi {\cal L}
+\partial_{q_i} \phi_z\,
\partial_{\phi_z} {\cal L}
+\partial_{q_i} \phi_t\,
\partial_{\phi_t} {\cal L}\big)
+\;\mbox{c.c.}\;\; 
\end{equation}
and
\begin{equation}\label{a8}
\partial_{\dot{q}_i} L
= \int_{-\infty}^{+\infty}\!\!\! d z \; 
\partial_{\dot{q}_i}\phi_t\,
\partial_{\phi_t} {\cal L}
+\;\mbox{c.c.}\;\;
\end{equation}
where c.c. stands for complex conjugate. Considering that $\phi$ is solution
of Eq.(\ref{s1.9}) (including the perturbative term $R[\phi]$), simple
manipulations allow one to obtain Lagrange-like equations for the $q_i$'s:
\begin{eqnarray}\label{a9}
\partial_{q_i} L-\frac{d}{dt}(\partial_{\dot{q}_i} L) & = &
\int_{-\infty}^{+\infty}\!\!\! d z \; 
\big[\partial_\phi{\cal L}-
\partial_z(\partial_{\phi_z}{\cal L})-
\partial_t(\partial_{\phi_t}{\cal L})
\big]\partial_{q_i}\phi+\;\mbox{c.c.} \nonumber\\
& = & 2\;\mbox{Re}\;\left\{\int_{-\infty}^{+\infty}\!\!\!dz \;
R^*[\phi_{\mbox{\tiny{sol}}}]\;
\partial_{q_i} \phi_{\mbox{\tiny{sol}}}\right\} \; .
\end{eqnarray}
In the particular case where
$\phi_{\mbox{\tiny{sol}}}(z,t) = \Phi(z-\bar{z}(t),\theta(t))$ one obtains
\begin{equation}\label{s2.5}
L(\theta,\dot{\theta},\bar{z},\dot{\bar{z}})=
\dot{\bar{z}}\,[\pi-2\,\theta-\sin(2\,\theta)]
-\frac{4}{3}\,\cos^3\theta\; ,
\end{equation}
and the equations of motion (\ref{a9}) read explicitly
\begin{equation}\label{s2.6.1}
4\,\dot{\theta}\,\cos^2\theta =
2\;\mbox{Re}\;\left\{\int_{-\infty}^{+\infty}\!\!\!dz \;
R^*[\phi_{\mbox{\tiny{sol}}}]\;
\partial_{\bar{z}} \phi_{\mbox{\tiny{sol}}}\right\} \; 
\end{equation}
\noindent and
\begin{equation}\label{s2.6.2}
4\,\cos^2\theta\,(\sin\theta-\dot{\bar{z}}) =
2\;\mbox{Re}\;\left\{\int_{-\infty}^{+\infty}\!\!\!dz \;
R^*[\phi_{\mbox{\tiny{sol}}}]\;
\partial_\theta \phi_{\mbox{\tiny{sol}}}\right\} \; .
\end{equation}
We note here a general feature, always valid in the framework of the adiabatic
approximation: with equation (\ref{s1.3}) conserving energy, one can show that the
soliton's energy defined in Eq.(\ref{a2}) has the same value far before
and far after the collision with the obstacle (the demonstration is
essentially the same as the one given in Section \ref{rad_ener} where, in
addition, the consequences of soliton's radiation -- neglected in the present
adiabatic approximation -- are taken into account). As a result, one obtains
for the solutions of Eqs(\ref{s2.6.1}) and (\ref{s2.6.2}) that
$\theta(+\infty)=\theta(-\infty)$, and $\dot{\bar{z}}(\pm\infty)= \sin
\theta(\pm\infty)=V$. Hence the soliton's shape and velocity may change
during the collision, but they eventually regain their initial
values. This is intimately connected to the neglecting of radiative effects in
the adiabatic approximation.

\section{}\label{calcul-ener-rad}
\setcounter{equation}{0} 

In this appendix we briefly indicate how to obtain expression (\ref{e7}) for
the radiated energy starting from Eq.(\ref{e6}), where $\delta\phi$ is given by
$\phi_1$-i.e., by (\ref{sd1}). Instead of giving a detailled
explanation on how to treat all the terms in the integrand of (\ref{e6}), for
brievity we focus on one of the contributions to the expression (\ref{e6}) for
$E_{\mbox{\tiny{rad}}}^\alpha$:
\begin{equation}\label{c2}
\int_{-\infty}^{+\infty}\!\!\!dx \;|\Phi|^2\,|\delta \phi^\alpha|^2
=
\int_{-\infty}^{+\infty}\!\!\!dx \;
\left(1-\frac{\cos^2\theta}{\cosh^2(x\cos\theta)}\right)
\, |\phi_1^\alpha(x,t)|^2 \; .
\end{equation}
We recall that we are interested of the evaluation of this term at large times.
Expressing $\phi_1$ through Eq.(\ref{sd1}), one can show that the term in
$\cosh^{-2}$ in the integrand on the RHS of Eq.(\ref{c2}) can be dropped
because it gives a contribution which decreases algebraically at $t\to +\infty$
(this can be checked by a stationary phase evaluation of the integrals over
the momenta). It thus remains to evaluate
\begin{equation}\label{c4}
\int_{-\infty}^{+\infty}\!\!\!dx \;|\phi_1^\alpha|^2=
\int_{-\infty}^{+\infty}\!\!\! d q \,
\int_{-\infty}^{+\infty}\!\!\! d p \,
\int_{-\infty}^{+\infty}\!\!\! d x \,\left\{
C_q^\alpha\,C_p^{\alpha\, *}
(u_q^\alpha u_p^{\alpha\,*}-v_q^{\alpha\,*} v_p^{\alpha\,*})
+C_q^{\alpha}C_p^{*\alpha}v_q^{\alpha} v_p^{*\alpha}
\right\}\; .
\end{equation}
In Eq.(\ref{c4}) we have added and substracted the contribution $v_q^{\alpha\,*}
v_p^{\alpha\,*}$ in order to make use of the normalization (\ref{b4}). For the
evaluation of the last part of the integrand on the RHS of Eq.(\ref{c4}), the
explicit expressions (\ref{b3}) of $u_q^\alpha(x)$ and $v_q^\alpha(x)$ are to
be used. In the course of this computation, an argument of stationary phase
shows that only the $x$-independent terms with $p=q$ give a finite contribution
at $t\rightarrow+\infty$. These terms will contribute as $2\pi\delta(p-q)$
after the integration over $x$. Altogether one obtains the expression
\begin{equation}\label{c5}
\int_{-\infty}^{+\infty}\!\!\!dx \;|\phi_1^\alpha|^2
\APPROX{t\to +\infty}
\int_{-\infty}^{+\infty}\!\!\! d q \,
|C_q^{\alpha}|^2\left\{{\cal N}_q^\alpha+
2\pi \left[ \left(\frac{q}{2} - \frac{\varepsilon_q^\alpha}{q}\right)^2
+ \cos^2\theta \right]^2\right\} \; .
\end{equation}
Noting that ${\cal N}_q^\alpha$ defined in Eqs(\ref{b4}) and (\ref{b5}) is an odd function
(and thus does not contribute to the integral since $|C_q^{\alpha}|^2$ is
even) and explicitly computing the other contributions, one obtains
\begin{equation}\label{c6}
\int_{-\infty}^{+\infty}\!\!\!dx \;|\phi_1^\alpha(x,t)|^2
\APPROX{t\to +\infty}8\,\pi
\int_0^{+\infty}\!\!\! d q \,
(q^2+2)
\left(\frac{\varepsilon_q^\alpha}{q}\right)^2
|D_q^\alpha(+\infty)|^2 \; .
\end{equation}

The others contributions to Eq.(\ref{e6}) can be computed similarly. One obtains
\begin{equation}\label{c7}
\int_{-\infty}^{+\infty}\!\!\!dx \;|\partial_x\phi_1^\alpha|^2
\APPROX{t\to +\infty}8\,\pi
\int_0^{+\infty}\!\!\! d q \,
q^2\,(q^2+2)
\left(\frac{\varepsilon_q^\alpha}{q}\right)^2
 |D_q^\alpha(+\infty)|^2 \; 
\end{equation}
\noindent and
\begin{equation}\label{c8}
\int_{-\infty}^{+\infty}\!\!\!dx \;(\Phi^* \phi_1^\alpha)^2
\APPROX{t\to +\infty} - 16\,\pi
\int_0^{+\infty}\!\!\! d q \, 
\left(\frac{\varepsilon_q^\alpha}{q}\right)^2 
|D_q^\alpha(+\infty)|^2 \; .
\end{equation}
Gathering all these contributions yields the final result (\ref{e7}).


\begin{thebibliography}{99}
\bibitem{r2} Nonstationary solutions do not fulfill Eq.~(\ref{s1.2}) for the
two following reasons: (i) at times when the perturbation of density reaches
$x\to\pm\infty$, the density of nonstationary solutions will not
exactly equal $n_\infty$ far from the obstacle, and (ii) an additional phase
difference (typically time dependent) will appear between
$x\to-\infty$ and $x\to+\infty$ (such as the one which occurs in presence of a
soliton).
\bibitem{r1} In dimensioned units, the velocity of sound in
an unperturbed system of constant density $n_\infty$ is
$c=(2\hbar\omega_\perp\,a_{sc} n_\infty/m)^{1/2}$.
\bibitem{trem} The fact that $\bar{z}$ depends on the fast time $t_0$ may seem
strange at first sight. This dependence is introduced in order to account for
its behavior in absence of obstacle: $\bar{z}=V\,t$.
\bibitem{att} This is not the only possible way of writing the solution of
Eq.(\ref{p7}). With the same degree of accuracy one can equivalently write
$\bar{z}=\int^{t_0}\sin[\theta(\epsilon\,t_0')]dt_0'+\hat{z}(t_1)$, where
	$\hat{z}$ is a (still unknown) function of $t_1$.
\bibitem{rem3} In dimensioned units this corresponds to a potential which
	varies slowly over a lengths scale of order of the healing length of the
	system.
\bibitem{temps} Note that $\tau$ is very close to the actual time $t$ and
	that, here, $t=0$ corresponds to the time where the soliton's center would be
	at $z=0$ in absence of the obstacle.
\end{thebibliography}
\end{document}